\documentclass[letterpaper,twocolumn,prl,aps,superscriptaddress,amsmath,amssymb,floatfix]{revtex4-1}
\usepackage{mathptmx}
\usepackage[latin9]{inputenc}
\usepackage{color}
\usepackage{amsmath}
\usepackage{amssymb}
\usepackage{graphicx}
\usepackage{esint}
\usepackage[unicode=true,
 bookmarks=true,bookmarksnumbered=false,bookmarksopen=false,
 breaklinks=false,pdfborder={0 0 1},backref=false,colorlinks=true]
 {hyperref}
\hypersetup{
 linkcolor=magenta,urlcolor=blue,citecolor=blue,pdfstartview={FitH},hyperfootnotes=false}

\makeatletter

\providecommand{\tabularnewline}{\\}

\usepackage{textcomp}
\usepackage{epstopdf}

\pdfpageheight\paperheight
\pdfpagewidth\paperwidth

\providecommand{\tabularnewline}{\\}

\@ifundefined{textcolor}{}{%
 \definecolor{BLACK}{gray}{0}
 \definecolor{WHITE}{gray}{1}
 \definecolor{RED}{rgb}{1,0,0}
 \definecolor{GREEN}{rgb}{0,1,0}
 \definecolor{BLUE}{rgb}{0,0,1}
 \definecolor{CYAN}{cmyk}{1,0,0,0}
 \definecolor{MAGENTA}{cmyk}{0,1,0,0}
 \definecolor{YELLOW}{cmyk}{0,0,1,0}
}

\usepackage{xcolor}\usepackage{soul}
\newcommand{\bra}[1]{\ensuremath{\left\langle#1\right|}}
\newcommand{\ket}[1]{\ensuremath{\left|#1\right\rangle}}

\definecolor{blue}{rgb}{0,0,1}
\definecolor{red}{rgb}{0,0,0}
\definecolor{green}{rgb}{0,1,0}
\newcommand{\red}[1]{\textcolor{red}{ #1}}

\makeatother

\begin{document}
\title{Bosonic quantum error correction codes in superconducting quantum
circuits}

\author{W.~Cai}
\thanks{These authors contributed equally to this work.}
\affiliation{Center for Quantum Information, Institute for Interdisciplinary Information
Sciences, Tsinghua University, Beijing 100084, China}
\author{Y.~Ma}
\thanks{These authors contributed equally to this work.}
\affiliation{Center for Quantum Information, Institute for Interdisciplinary Information
Sciences, Tsinghua University, Beijing 100084, China}
\author{W.~Wang}
\thanks{These authors contributed equally to this work.}
\affiliation{Center for Quantum Information, Institute for Interdisciplinary Information
Sciences, Tsinghua University, Beijing 100084, China}
\author{C.-L.~Zou}
\email{clzou321@ustc.edu.cn}
\affiliation{Key Laboratory of Quantum Information, CAS, University of Science
and Technology of China, Hefei, Anhui 230026, P. R. China}
\author{L.~Sun}
\email{luyansun@tsinghua.edu.cn}
\affiliation{Center for Quantum Information, Institute for Interdisciplinary Information
Sciences, Tsinghua University, Beijing 100084, China}

\begin{abstract}
\textbf{Quantum information is vulnerable to environmental noise
and experimental imperfections, hindering the reliability of practical
quantum information processors. Therefore, quantum error correction
(QEC) \red{that can protect quantum information against noise} is vital for universal and scalable quantum computation. Among
many different experimental platforms, superconducting quantum
circuits and bosonic encodings in superconducting microwave modes
are appealing for their unprecedented potential in QEC. During the
last few years, bosonic QEC is demonstrated to reach the break-even
point, i.e. the lifetime of a logical qubit is enhanced to exceed that
of any individual components composing the experimental system. \red{Beyond
that, universal gate sets and fault-tolerant operations on the bosonic codes are also realized}, pushing quantum information processing towards the QEC era. In this article, we review the recent progress of the bosonic codes, including the Gottesman-Kitaev-Preskill codes, cat codes, and binomial codes, and discuss the opportunities of bosonic codes in various quantum applications, ranging from fault-tolerant quantum computation to quantum metrology. We also summarize the challenges \red{associated with the bosonic codes} and provide an outlook for the potential research directions in the long terms.}

\end{abstract}
\maketitle

\section{Introduction}
\label{sec:introduction}
Quantum computers promise to exponentially or dramatically outperform classical computers on certain problems (e.g. factoring and unstructured database searching) because of quantum coherence and true parallel computation~\cite{Preskill1998,Nielsen2000,Devoret2013,Preskill2018}. However, quantum states are fragile and can be easily destroyed by their inevitable coupling to the uncontrolled environment, which presents a major obstacle to universal quantum computation~\cite{Nielsen2000,Cho2020}. A practical quantum computer that is capable of large circuit depth, therefore, ultimately calls for operations on logical qubits protected by quantum error correction (QEC) against unwanted or uncontrolled errors and is expected to spend a vast majority of its resources on error correction~\cite{Shor1995,Steane1996,Gottsman2010,Fowler2012,Devitt2013,Roffe2019}. The realization of such a logical qubit with a longer coherence time than its individual physical components is considered as one of the most
challenging and urgent goals for current quantum information processing~\cite{Devoret2013,Ofek2016}. \red{When universal gate sets on these logical qubits are available, quantum information processing technologies would enter an era of quantum protection. Finally, universal quantum computation is realizable by scaling up the system when the error rates of the logical gates exceed a certain threshold.}


Extensive attention has been paid to the qubit-based quantum computation systems. However, the demonstration of QEC in those systems is extremely challenging due to the huge physical resource overhead and the difficulties in scaling up the number of qubits~\cite{Fowler2012,Gidney2019,Corcoles2020}. So far, qubit-based QEC encoding and gate operations on the encoded qubits still remain elusive. Compared with qubits, harmonic oscillators or bosonic modes provide an alternative route towards universal quantum computation. The bosonic modes are beneficial to quantum information processing in four aspects. First, a single bosonic mode can provide an infinitely large Hilbert space, which allows QEC encoding by only extending excitation numbers while keeping the noise channels fixed. Second, bosonic modes could be realized with multiple degrees of freedom, e.g. spatial, temporal, frequency, polarization, or their combinations, and thus are scalable. Third, bosonic modes are convenient in transferring information and also can easily interface with many different physical systems, therefore they are inevitable building blocks in quantum networks. Lastly, bosonic modes are fundamental and indispensable physical systems that cannot be replaced by other qubit or finite-level quantum systems.

\begin{figure*}
\centering{}\includegraphics[width=1.90\columnwidth]{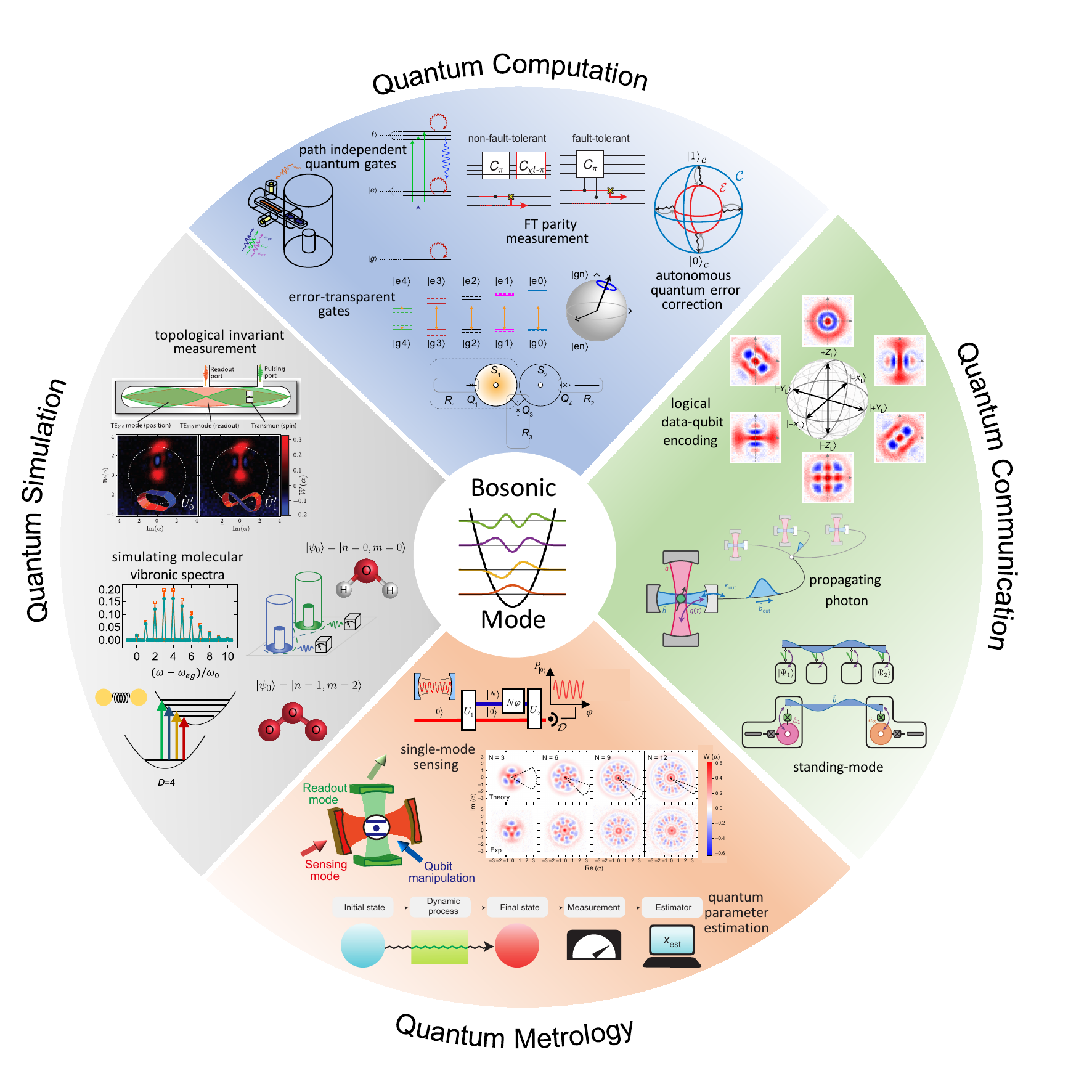}
\caption{\textbf{Quantum applications of bosonic modes.} Bosonic modes have
wide applications in quantum computation, quantum communication, quantum
simulation, and quantum metrology. Here we only list a small portion of them.
For quantum computation, QEC and fault-tolerant operations on bosonic
codes have been demonstrated. Adapted from Refs.~\cite{Rosenblum_2018,reinhold2020errorcorrected,Ma_2020,Gertler2020}.
For quantum communication, quantum state transfer, remote
entanglement, gate teleportation, etc., have been demonstrated. Adapted from Refs.~\cite{Chou2018,burkhart2020errordetected}.
A single bosonic mode can be employed for quantum metrology to achieve
a measurement precision beyond the shot-noise limit. Moreover, it
is promising to achieve quantum-enhanced sensing by constructing suitable
QEC codes. Adapted from Ref.~\cite{WangNC2019Heisenberg,Escher2011}.
Bosonic modes also can be used to simulate solid-state materials and
molecular vibrations. Adapted from Refs.~\cite{Flurin2017,Hu2018simulation,WangChristopher2020}.}
\label{fig:BosonicCodesApplications}
\end{figure*}

Therefore, bosonic modes have attracted a lot of interest in quantum information processing and demonstrated wide applications in quantum computation, quantum communication, quantum simulation, and quantum metrology in the last decades, as shown in Fig.~\ref{fig:BosonicCodesApplications}. Bosonic modes with QEC protection would exhibit better quantum properties and thus will greatly broaden the above applications. \red{As a result, extensive explorations of QEC based on bosonic modes are demanded.} As mentioned, QEC in a bosonic mode benefits from the infinite-dimensional Hilbert space of a harmonic oscillator for redundant information encoding and only one error syndrome that needs to be monitored, thus greatly reducing the requirements on hardware. The bosonic modes could be realized with microwave or optical photons, magnons, phonons, plasmons, as well as polaritons~\cite{Rivera2020}. Among them, superconducting circuit quantum electrodynamics (circuit QED) architecture~\cite{Blais2004,Wallraff2004,You2011,Devoret2013,Gu2017,BlaisNP2020,Blais2020cQED} is of particular interest for bosonic QEC codes due to its unprecedented capability in quantum control. Three dimensional (3D) cavities~\cite{Paik2011}, especially 3D coaxial cavities~\cite{ReagorAPL2013}, exhibit great quantum coherence with single-photon lifetimes up to 1-10~ms. In analogy to optical cavity QED that studies the interaction between atoms and photons, circuit QED describes the interaction between superconducting qubits (artificial atoms) and microwave photons in a cavity with ultrahigh cooperativities. Thus, this experimental platform allows universal control of the bosonic mode with high fidelities, and QEC that exceeds or closely reaches the break-even point~\cite{Ofek2016,Hu2019,campagneibarcq2020}, logical-qubit operations~\cite{Heeres2017,Chou2018,Hu2019,XuPRL2020PhotonicQubits,campagneibarcq2020}, and fault-tolerant operations~\cite{Rosenblum_2018,reinhold2020errorcorrected,Ma_2020} have already been demonstrated based on bosonic codes.

This article reviews the recent development of bosonic QEC codes in superconducting quantum circuits. The organization is as follows. A basic introduction of QEC and bosonic modes is provided in Sec.~\ref{sec:QECschemes}. Details on the three most widely used bosonic codes, i.e. the Gottesman-Kitaev-Preskill (GKP), cat, and binomial codes are presented in Sec.~\ref{sec:BosonicCodes}. In Sec.~\ref{sec:UniversalControl}, we summarize the underlying kernel techniques to realize the bosonic codes as well as the universal control of the codes. With the basic toolkit available, intriguing potential applications of bosonic codes and their proof-of-principle demonstrations in the fault-tolerant quantum computation, quantum communication, quantum simulation, and quantum metrology are presented in Sec.~\ref{sec:Applications}. Finally, future directions and challenges are discussed in Sec.~\ref{sec:discussions}.

\section{Basics of QEC}
\label{sec:QECschemes}
\begin{figure}
\centering{}\includegraphics[width=1\columnwidth]{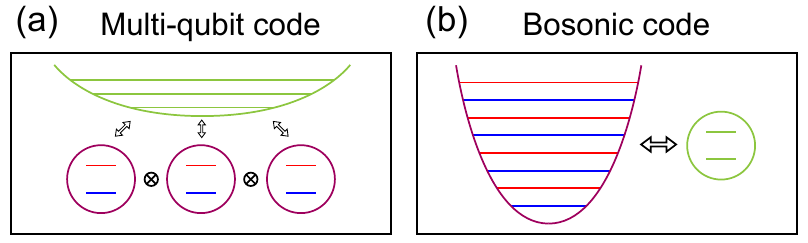}
\caption{\textbf{Multi-qubit architecture vs bosonic-mode architecture.} Enlarged
Hilbert space can be constructed with multiple qubits (a) or one bosonic mode (b). Qubits and harmonic oscillators are critical elements for these two different architectures. However, their roles are exchanged: in the qubit-based architecture, quantum information is stored on the qubits while harmonic oscillators are used to couple or readout the qubits; in the bosonic architecture, quantum information is stored in the bosonic modes while the qubits provide the necessary nonlinearity for the control and readout of the bosonic modes.}
\label{fig:multiqubitVsBosonicCodes}
\end{figure}

In this section, we provide a brief introduction of the QEC codes and the basic properties of a bosonic mode. For more detailed discussions on QEC and fault-tolerance in the context of the qubit model, \red{we suggest the review articles Refs.~\cite{Gottsman2010,Devitt2013,Terhal2015,Campbell2017,Roffe2019} for further reading. We also refer the readers to Refs.~\cite{Terhal_2020,Joshi2020} for related reviews on bosonic codes.}

The key idea of classical error correction to protect information against noise is to encode the information with added redundancy. By doing so, even if some information in the encoded message is corrupted by noise, there is still enough redundancy in the encoded information to fully recover the original information. For example, the classical repetition codes are to use odd multiples of 0's and 1's to represent the logical 0 and logical 1 respectively. The only classical error of bit-flips can be corrected by the majority voting, and this type of error-correcting codes can suppress the leading orders of errors.



Similar to the code redundancy in the classical error correction, QEC is possible by expanding the Hilbert space of a logical qubit~\cite{Shor1995,Steane1996}. Different from the classical case, quantum information could be any superposition of codewords that occupies a subspace of the expanded Hilbert space, called the code space. Restricted by quantum coherence, QEC cannot measure the codewords directly, but rather measure the so-called error syndromes to diagnose possible errors without perturbing the encoded information. By constructing QEC codes, these requirements could be satisfied if errors due to noise could turn the code space into different error spaces. Then, errors that have occurred could be detected by distinguishing different subspaces, and appropriate recovery operations can be applied to restore the original quantum information by mapping the error space back to the code space.

For example, the QEC codes could be constructed with multiple qubits, as illustrated in Fig.~\ref{fig:multiqubitVsBosonicCodes}a. Quantum information is encoded with a simple repetition
code $\mathrm{span}\left\{ \left|000\right\rangle ,\left|111\right\rangle \right\} $. In such a way, quantum information is essentially stored non-locally through entanglement among the physical
qubits, while single physical qubits contain no encoded information. Because noise is generally local and independent, it only corrupts little about the stored information. A bit-flip error on the code could be detected by measuring the correlations between neighboring qubits (the error syndromes) instead of projective measurements of the encoded quantum states. The essential part for QEC to work is that these error syndrome measurements need to be non-destructive to the encoded information, which is realized by introducing and measuring ancillary qubits or modes that interact with the physical qubits constituting the QEC codewords.

The above QEC would be properly described by a more general mathematical framework. The QEC condition is a sufficient and necessary
condition for a QEC code to protect against errors in a given error set $\varepsilon=\{\hat{E_{i}}\}$:
\begin{equation}
\begin{aligned}\hat{\mathcal{P}}\hat{E_{i}}^{\dagger}\hat{E_{j}}\hat{\mathcal{P}}=\alpha_{ij}\hat{\mathcal{P}}\end{aligned}
\label{eq:QECconditions}
\end{equation}
where $\hat{\mathcal{P}}$ is the projection operator onto the code space and $\alpha_{ij}$ is a Hermitian matrix~\cite{Nielsen2000}. Consequently, we have:
\begin{equation}
\begin{aligned}\bra{0_{L}}\hat{E_{i}}^{\dagger}\hat{E_{j}}\ket{0_{L}}=\bra{1_{L}}\hat{E_{i}}^{\dagger}\hat{E_{j}}\ket{1_{L}},\end{aligned}
\label{eq:QEClogicalstateindis}
\end{equation}
and
\begin{equation}
\begin{aligned}\bra{0_{L}}\hat{E_{i}}^{\dagger}\hat{E_{j}}\ket{1_{L}}=\bra{1_{L}}\hat{E_{i}}^{\dagger}\hat{E_{j}}\ket{0_{L}}=0,\end{aligned}
\label{eq:QEClogicalstateOrthognonal}
\end{equation}
where $\ket{0_{L}}$ and $\ket{1_{L}}$ are the basis states of the codewords. Equation~\ref{eq:QEClogicalstateindis} requires that the logical states are indistinguishable under different errors, independent of the codewords. Otherwise, the codewords will suffer the deformation error and the environment could potentially distinguish these two basis states and eventually induces uncorrectable errors. Equation~\ref{eq:QEClogicalstateOrthognonal} requires that all the spaces are orthogonal to each other.

In the past decades, most of the theoretical and experimental efforts are spent on the qubit-based QEC codes. The concatenated encoding is proposed for fault-tolerance, but it is tremendously challenging because
of the required extremely low gate error threshold and large resource overhead~\cite{Shor1995,Steane1996}. The recently developed surface codes~\cite{Fowler2012}, which use the topological property of a large number of qubits in a two-dimensional grid to protect against external noise, can tolerate a much higher error rate $\sim1\%$, but still at the cost of huge resource overhead. Both of these approaches need to scale up the number of physical qubits for achieving QEC. There is a lot of experimental progress along this line, for example in trapped-ion systems~\cite{Schindler2011,Nigg2014}, nitrogen-vacancy centers in diamond~\cite{TaminiauNatureNanoTech2014,Cramer2016}, and superconducting circuits~\cite{Reed2012,Kelly2015}. However, to have extended lifetime than the physical qubits and to have logical operations are difficult to achieve with the multi-qubit encoding because the number of distinct error channels increases with the number of qubits, and non-local gates on a collection of physical qubits are required.

\begin{figure*}
\centering{}\includegraphics[width=1.8\columnwidth]{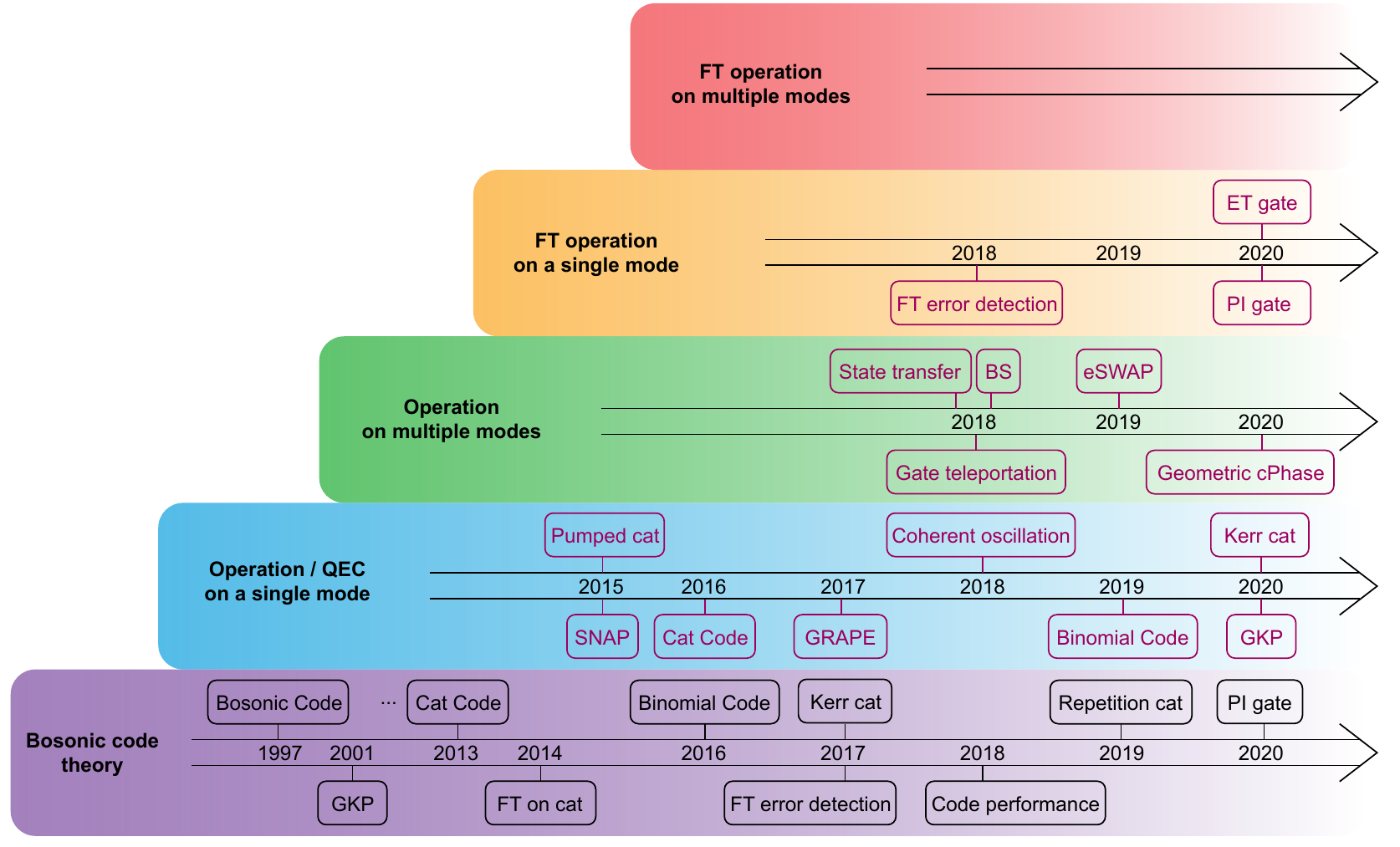}
\caption{\textbf{Roadmap of bosonic codes.} Tremendous progress of the bosonic codes has been made in the last two decades. Theories of bosonic codes as the foundation are listed at the bottom~\cite{Chuang1997,GottesmanPRA2001,Leghtas_2013,Mirrahimi_2014,Michael2016,Cohen_2017,PuriNPJ2017,AlbertPRA2018,PuriPRX2019,GuillaudPRX2019,ma2019pathindependent}.
The above rows list representative steps in the experimental development
of bosonic codes in superconducting circuit QED architectures. QEC
and gate operations have been demonstrated based on cat codes~\cite{Ofek2016},
binomial codes~\cite{Hu2019}, and GKP codes~\cite{campagneibarcq2020}.
Photon-number-selective arbitrary phase (SNAP) gate~\cite{HeeresPRL2015}
and optimal control technique based on gradient ascent pulse engineering
(GRAPE)~\cite{Heeres2017} are important for universal control of
the bosonic mode. Pumped and Kerr cat qubits~\cite{Leghtas_2015,Touzard2018PRX,GrimmKerrcat2020}
have been developed with biased noise for potentially important QEC
applications. Operations on multiple modes such as state transfer~\cite{Axline2017},
gate teleportation~\cite{Chou2018}, beam-splitter (BS) interaction~\cite{GaoPRX2018},
exponential SWAP gate~\cite{GaoNature2019}, and geometric controlled-phase
gate~\cite{XuPRL2020PhotonicQubits} have also been demonstrated.
Fault-tolerant (FT) error detection~\cite{Rosenblum_2018}, path-independent
(PI) phase gate~\cite{reinhold2020errorcorrected}, and error-transparent
(ET) gate~\cite{Ma_2020} on a single bosonic mode have been developed.
\red{In the future, FT control needs to be extended to multiple modes for universal quantum information processing.}}
\label{fig:BosonicCodesProgress}
\end{figure*}

\begin{table*}[t]
\vspace{12bp}
\begin{centering}
\begin{tabular*}{0.9\textwidth}{@{\extracolsep{\fill}}@{\extracolsep{\fill}}@{\extracolsep{\fill}}@{\extracolsep{\fill}}ccccccc}
\hline
Year & Code & Ancilla $T_{1}$ & Ancilla $T_{2}^{*}$ & Fock $\{\ket{0},\ket{1}\}$ encoding & Uncorrected code & Corrected code\tabularnewline
\hline
2016~\cite{Ofek2016} & Cat & $35\ \mu$s & $12\ \mu$s & $287\ \mu$s & $147\ \mu$s & $318\ \mu$s\tabularnewline
2019~\cite{Hu2019} & Binomial & $30\ \mu$s & $40\ \mu$s & $216\ \mu$s & $71\ \mu$s & $200\ \mu$s\tabularnewline
2020~\cite{Ma_2020} & Binomial (ET) & $35\ \mu$s & $25\ \mu$s & - & $185\ \mu$s & $364\ \mu$s\tabularnewline
2020~\cite{campagneibarcq2020} & GKP (Square) & $50\ \mu$s & $60\ \mu$s & $245\ \mu$s ($T_{1}$) & - & $275\ \mu$s ($XZ$) $160\ \mu$s ($Y$)\tabularnewline
2020~\cite{campagneibarcq2020} & GKP (Hex) & $50\ \mu$s & $60\ \mu$s & $245\ \mu$s ($T_{1}$) & - & $205\ \mu$s\tabularnewline
2020~\cite{Gertler2020} & Cat (AQEC) & $39\ \mu$s & $17\ \mu$s & $440\ \mu$s & $130\ \mu$s & $288\ \mu$s\tabularnewline
\hline
\end{tabular*}
\par\end{centering}
\caption{\textbf{Experimental performance of various bosonic codes.} The right
three columns list the measured process fidelity decay times except
for the GKP experiment (state decay times).}
\label{Table:coherenttime}
\end{table*}

Alternatively, there is another strategy which uses a single quantum system with intrinsically large Hilbert space, instead of a collection of physical qubits, to redundantly encode quantum information and would significantly reduce the required hardware. One can realize QEC first and then scale up for more complicated quantum information processing applications. \red{A harmonic oscillator or a bosonic mode is just such a system that supports an infinitely large Hilbert space of Fock states and has long been proposed to store quantum information.} Utilizing the redundancy of the Hilbert space, QEC could be constructed in a single bosonic mode. The main advantage of such a QEC scheme is that the large Hilbert space is achieved in only a single degree of freedom so that the associated errors are restricted.

The basic bosonic code architecture is shown in Fig.~\ref{fig:multiqubitVsBosonicCodes}b. A coupled non-linear element, typically a two-level qubit, is also essential for arbitrarily controlling and manipulating quantum states of the harmonic oscillator. Details about the universal control of this composite system will be discussed in Sec.~\ref{sec:UniversalControl}. \red{In this article, we will only focus on microwave photons in superconducting microwave cavities which are excellent harmonic oscillators with long lifetimes (up to 1-10~ms~\cite{ReagorAPL2013}) and ideal for quantum memories or logical qubits in the first place.}

For a simple case with the error set $\varepsilon=\{\hat{I},\hat{a}\}$, where $\hat{a}$ denotes the single-photon-loss error due to the damping, there is obviously only one error syndrome, photon number parity or generalized parity, that needs to be monitored continuously. To meet the QEC condition Eq.~\ref{eq:QEClogicalstateindis}, the basis states of the logical qubit should satisfy:
\begin{equation}
\begin{aligned}
\bra{0_{L}}\hat{a}^{\dagger}\hat{a}\ket{0_{L}}=\bra{1_{L}}\hat{a}^{\dagger}\hat{a}\ket{1_{L}}.
\end{aligned}
\end{equation}
This requires that the codewords should have equal average photon numbers. It can also be understood as the environment should not distinguish the two logical basis states from a photon loss event in order to preserve the encoded quantum information. Note that for continuous damping or attenuation the above error set $\{\hat{I},\hat{a}\}$ is only approximate and not bounded \red{(cannot be normalized to satisfy the condition $\sum_{i}\hat{E_{i}}^{\dagger}\hat{E_{i}}=\hat{I}$)}. The exact expression of the error set for photon loss errors is:
\begin{equation}
\begin{aligned}
\hat{E_{l}}=\sqrt{\frac{(1-e^{-\eta})^{l}}{l!}}e^{-\frac{\eta}{2}\hat{a}^{\dagger}\hat{a}}\hat{a}^{l}.
\label{eq:ExactPhotonLossError}
\end{aligned}
\end{equation}
where \red{$l=0,1,...$ and $\eta=\kappa t\ll1$ is the photon loss coefficient for storing quantum information with a mode dissipation rate $\kappa$ and a duration $t$. We can also have $\eta=\alpha d$ for transmitting photons over a distance $d$ with a channel attenuation coefficient $\alpha$~\cite{Li2017}.}

We finally note that when more photons are added to the microwave cavity for information redundancy, although no more type of errors is introduced, the error rate becomes $n$ times larger ($n$ is the average photon number in the codewords). This is the price one has to pay for any QEC schemes: redundantly encoding quantum information always increases either the number of error channels or the error rate. However, good control and QEC on the logical qubits hopefully can compensate this negative effect and eventually lead to better protection of quantum information with extended coherence.

\begin{figure*}
\centering{}\includegraphics[width=1.8\columnwidth]{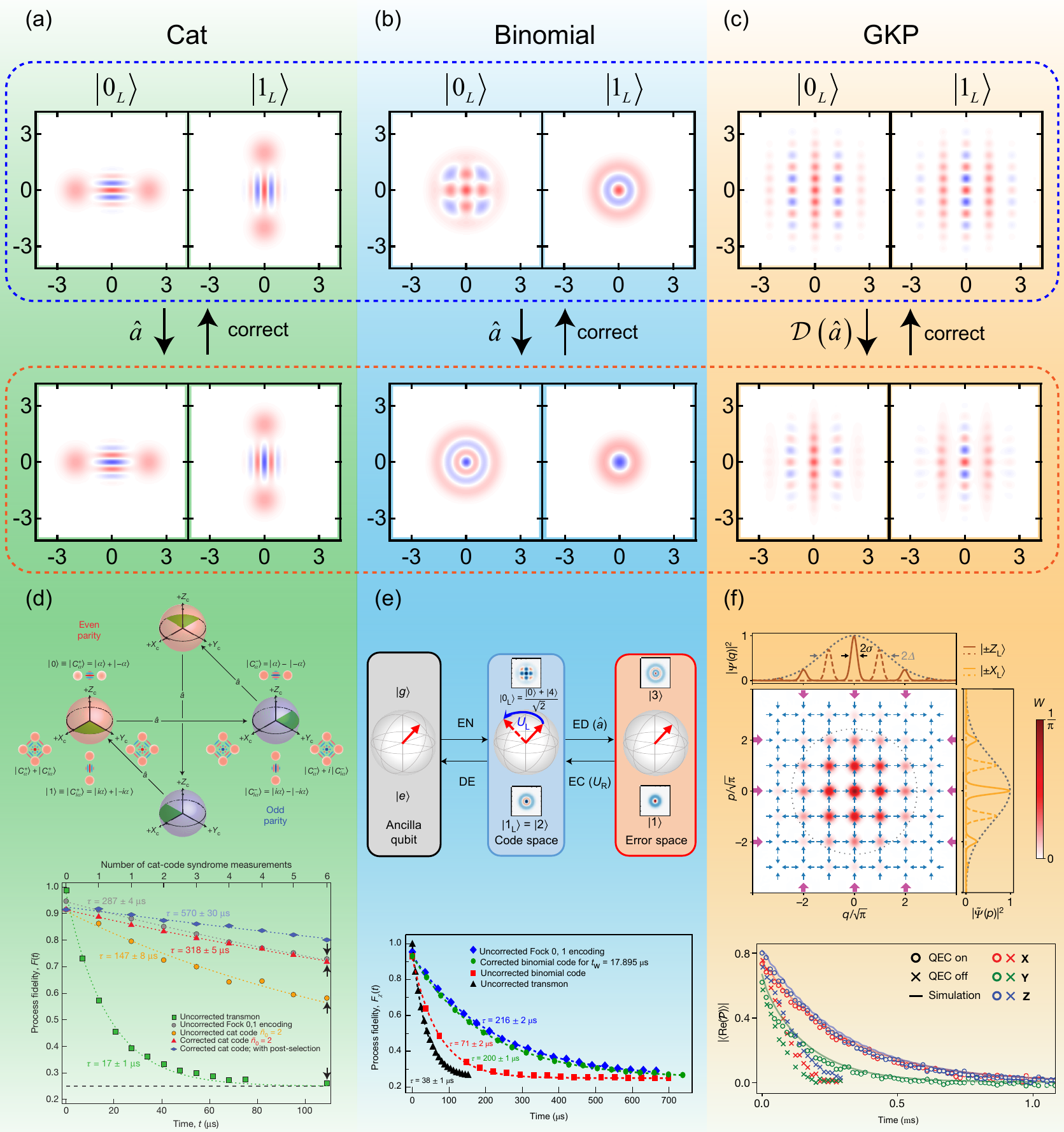}
\caption{\textbf{QEC with three typical bosonic codes.} (a-c) Wigner functions of the logical qubit basis states in the code and error spaces for the cat, binomial,
and GKP codes, respectively. (d-f) Experimental demonstration of QEC based on these three bosonic codes. Adapted from Refs.~\cite{Ofek2016,Hu2019,campagneibarcq2020}.}
\label{fig:QEC_CatBinomialGKP}
\end{figure*}



\section{QEC based on bosonic codes}
\label{sec:BosonicCodes}
Figure~\ref{fig:BosonicCodesProgress} summarizes the recent progress of the bosonic codes including both theoretical and experimental developments. \red{Here, we concentrate on three bosonic codes based on a single bosonic mode, i.e. GKP, cat, and binomial codes, and provide discussions on the related theories and the recent experimental progress. The experimental achievements of these three types of codes are summarized in Fig.~\ref{fig:QEC_CatBinomialGKP} and Table~\ref{Table:coherenttime}.}


\subsection{Cat codes}
\label{sec:CatCodes}
Coherent states are quasi-classical states that can be readily generated with classical methods, and hence they have been widely used in communication. Because phase is more robust against photon loss error, information is typically encoded in the phase of a coherent state. In analogy to classical phase-shift keying, quantum information can also be encoded to the phase of a coherent state. The simplest code (two-component cat code) is thus to use two coherent states with opposite phases for its two basis states:
\begin{equation}
\begin{aligned}\ket{0} & =\ket{\alpha},\\
\ket{1} & =\ket{-\alpha}.
\end{aligned}
\label{eq:2legCat}
\end{equation}
However, this code does not have redundancy and is not error correctable, because when the single-photon-loss error $\hat{a}$ occurs the state remains in the same code space.

The four-component cat code is later proposed to encode quantum information in a superposition state of coherent states with four different phases~\cite{Leghtas_2013}, as shown in Fig.~\ref{fig:QEC_CatBinomialGKP}a. With the extra degrees of freedom, this code has the necessary redundancy to fight against photon loss errors: two dimensions $\{\ket{0_{L}},\ket{1_{L}}\}$ for the encoding, while the other two $\{\ket{0_{E}},\ket{1_{E}}\}$
for error detection. The code basis states and error basis states are respectively:
\begin{equation}
\begin{aligned}\ket{0_{L}} & =C_{\alpha}^{+}=\mathcal{N}(\ket{\alpha}+\ket{-\alpha}),\\
\ket{1_{L}} & =C_{i\alpha}^{+}=\mathcal{N}(\ket{i\alpha}+\ket{-i\alpha}),
\end{aligned}
\label{eq:4legCat_codespace1}
\end{equation}
and
\begin{equation}
\begin{aligned}\ket{0_{E}} & =C_{\alpha}^{-}=\mathcal{N}(\ket{\alpha}-\ket{-\alpha}),\\
\ket{1_{E}} & =C_{i\alpha}^{-}=\mathcal{N}(\ket{i\alpha}-\ket{-i\alpha}),
\end{aligned}
\label{eq:4legCat_errorspace1}
\end{equation}
where $\mathcal{N}\approx1/\sqrt{2}$ is the normalization factor.
The code basis states have the same average photon number, but are
not exactly orthogonal as preferred for information encoding unless
$\alpha$ is sufficiently large.


To avoid the non-orthogonality error, the above cat codes can be made
orthogonal by using states with well-defined generalized parities:
\begin{equation}
\begin{aligned}\ket{0_{L}^{\prime}} & =\mathcal{N}_{0}(C_{\alpha}^{+}+C_{i\alpha}^{+})=\mathcal{N}_{0}(\ket{\alpha}+\ket{-\alpha}+\ket{i\alpha}+\ket{-i\alpha}),\\
\ket{1_{L}^{\prime}} & =\mathcal{N}_{0}(C_{\alpha}^{+}-C_{i\alpha}^{+})=\mathcal{N}_{0}(\ket{\alpha}+\ket{-\alpha}-\ket{i\alpha}-\ket{-i\alpha}),
\end{aligned}
\label{eq:4legCat_codespace2}
\end{equation}
where $\mathcal{N}_{0}\approx1/2$. $\ket{0_{L}^{\prime}}$ contains photon number states that are multiples of four, while $\ket{1_{L}^{\prime}}$ contains photon number states that are even but not multiples of four. However, these two basis states do not have the same average photon numbers unless $\alpha$ is large enough or at certain sweet spots.

Therefore, at large enough $\alpha$ both of these two codes satisfy
the QEC conditions (Eqs.~\ref{eq:QEClogicalstateindis} and \ref{eq:QEClogicalstateOrthognonal}),
and can both efficiently correct the single-photon-loss error in the error set $\varepsilon=\{\hat{I},\hat{a}\}$. For both codes, the code and error spaces have exact photon-number parities of even and odd, respectively. Photon parity is thus the error syndrome for error detection, which can be readily realized in a quantum non-demolition manner in a circuit QED architecture~\cite{SunNature2014}.

These codes have the following two major properties. First, single-photon-loss errors cause quantum jumps of the encoded state between the code and error spaces and each error is accompanied by a phase shift of $\pi/2$ about the $Z$ axis within the logical space, but without corrupting the encoded quantum information. The original information is not fully recovered until after the fourth photon loss error. Remarkably, this property makes explicit error correction after each error detection unnecessary and only requires tracking of the number of errors and implementing the recovery operation at the end of the whole QEC process. Therefore, the deleterious effect of imperfect recovery operations can be eliminated.





Second, in the absence of quantum jumps the quantum state deterministically shrinks towards the vacuum state. This inevitable property demands one to re-pump energy into the codeword before the coherent states start to overlap
and cause a considerable non-orthogonality error. \red{However, this error could not be fully corrected by a unitary operation.}

To mitigate the non-orthogonality error, there are two strategies to continuously
pump or stabilize the cat codes. One requires four-photon driven dissipative process with specifically engineered four-photon dissipation~\cite{Mirrahimi_2014}:
\begin{equation}
\begin{aligned}\frac{d\rho}{dt}=\mathcal{L}[\sqrt{\kappa_{4ph}}(\hat{a}^{4}-\alpha^{4})]\rho,
\label{eq:4photondissipation}\end{aligned}
\end{equation}
where $\mathcal{L}$ is the Lindblad superoperator and $\kappa_{4ph}$ is the four-photon dissipation rate. As a result, the four states $\ket{\pm\alpha}$ and $\ket{\pm i\alpha}$ are the steady states and the logical states will be confined in this manifold.

The other strategy that can achieve the same stabilization is to engineer a specific Hamiltonian~\cite{PuriNPJ2017}:
\begin{equation}
\begin{aligned}\hat{H}=-K\hat{a}^{\dagger4}\hat{a}^{4}+\epsilon_{4}(\hat{a}^{\dagger4}+\hat{a}^{4}),\label{eq:4photonKerrHamiltonian}\end{aligned}
\end{equation}
where $K$ is the coefficient of the high-order Kerr non-linearity and $\epsilon_{4}$ is the amplitude of the four-photon drive. Then the four states $\ket{\pm\alpha}$ and $\ket{\pm i\alpha}$ are the eigenstates of this Hamiltonian, and the adiabatic theorem ensures the logical states to be confined in this manifold. By tracking the photon number parity, the dynamics can be restricted to the even parity states (or the code space).

When the cat codes are stabilized by one of the above strategies, fault-tolerant gates can be realized through a two-photon drive $\hat{a}^{\dagger2}+\hat{a}^{2}$ for arbitrary rotations around $X$ and a beam-splitter-like drive $\hat{a}_{1}^{\dagger2}\hat{a}_{2}^{2}+\hat{a}_{2}^{\dagger2}\hat{a}_{1}^{2}$ for the two-qubit entangling gate~\cite{Mirrahimi_2014}. Both strategies share similar principle and face difficulties of relatively strong four-photon drives~\cite{Mundhada_2019}, awaiting experimental realization.

Cat codes can tolerate more errors by increasing the number of coherent state components. The basis states of a $n$-component cat state can be written as~\cite{Li2017}
\begin{equation}
\begin{aligned}\ket{0_{L}} & \propto\sum_{k=1}^{n}\ket{\alpha e^{i2k\pi/n}},\\
\ket{1_{L}} & \propto\sum_{k=1}^{n}e^{i4k\pi/n}\ket{\alpha e^{i2k\pi/n}}.
\end{aligned}
\end{equation}
This code can correct photon loss errors up to $n/2-1$ order. But the average photon number should be increased to satisfy the orthogonality condition required by QEC. \red{Cat codes can also be extended to multiple modes that can protect against photon loss via either active syndrome measurement or an autonomous procedure. For example, the pair-cat codes~\cite{Albert_2019} occupying two modes can protect against arbitrary number of photon loss errors in one mode given the other one has no error.}


Experimentally, the cat code (Eq.~\ref{eq:4legCat_codespace1}) is the first bosonic code that surpasses the break-even point~\cite{Ofek2016}, as shown in Fig.~\ref{fig:QEC_CatBinomialGKP}d, benefiting from its special property that tracking the number of errors without immediate recover operation is equivalent to having corrected the state. Universal control of a logical qubit with the cat coding (Eq.~\ref{eq:4legCat_codespace1}) has been realized separately by numerically optimized pulses~\cite{Heeres2017} (see Sec.~\ref{sec:UniversalControl} for more details). A controlled-phase (cPhase) gate between two coherent-state encodings (Eq.~\ref{eq:2legCat}) has also been realized~\cite{XuPRL2020PhotonicQubits}.

Lastly, we want to mention that simpler ideas than Eqs.~\ref{eq:4photondissipation}
and \ref{eq:4photonKerrHamiltonian} have been experimentally realized for a two-photon drive case~\cite{Leghtas_2015,Touzard2018PRX,Lescanne2020,GrimmKerrcat2020} based on the two similar strategies:
\begin{equation}
\begin{aligned} & \frac{d\rho}{dt}=\mathcal{L}[\sqrt{\kappa_{2ph}}(\hat{a}^{2}-\alpha^{2})]\rho\end{aligned}
\end{equation}
and
\begin{equation}
\begin{aligned} & \hat{H}=-K\hat{a}^{\dagger2}\hat{a}^{2}+\epsilon_{2}(\hat{a}^{\dagger2}+\hat{a}^{2}).\end{aligned}
\end{equation}
In these two cases, the stabilized manifold is $\{\ket{\alpha},\ket{-\alpha}\}$. Since this Hilbert space is not large enough for error correction, it only defines the so-called cat qubit.

Although this type of qubit cannot be protected against photon loss error, it has a very special and interesting property, i.e., its noise is biased. This can be understood for the cat qubit defined as:
\begin{equation}
\begin{aligned}\ket{0}_{\alpha} & =\frac{1}{\sqrt{2}}(C_{\alpha}^{+}+C_{\alpha}^{-})=\ket{\alpha}+O(e^{-2|\alpha|^{2}}),\\
\ket{1}_{\alpha} & =\frac{1}{\sqrt{2}}(C_{\alpha}^{+}-C_{\alpha}^{-})=\ket{-\alpha}+O(e^{-2|\alpha|^{2}}).
\end{aligned}
\label{eq:catqubit}
\end{equation}
The built-in stabilization mechanism provides a natural protection of this qubit. The encoded information is non-local in the phase space of the harmonic oscillator. The distance between the two basis states
thus prevents any noise process that induces local displacement in phase space. As a result, the bit-flip error is exponentially suppressed with the average number of photons, while the phase-flip error only
increases linearly. Therefore, the cat qubit is noise biased and this biased structure of noise has been observed experimentally~\cite{Lescanne2020,GrimmKerrcat2020}. Coherent rotations around $X$ on such a stabilized cat qubit have also been demonstrated~\cite{Touzard2018PRX}.

Furthermore, due to the infinite-dimensional Hilbert space that the cat qubits are embedded in, a universal set of bias-preserving gates can be realized on the cat qubits~\cite{PuriBiasPreservingGates2019,GuillaudPRX2019},
which however is not possible for regular two-level systems. Fault-tolerant error syndrome detection can also be achieved based on the biased-noise cat qubit as the ancilla~\cite{PuriPRX2019}. The cat qubits can also be the promising building blocks for a surface code tailored to biased noise with high error thresholds~\cite{TuckettPRL2018,TuckettPRX2019,Tuckett2020}.
Therefore, the biased-noise cat qubits are important resources for fault-tolerant quantum computation.

\begin{figure*}
\centering{}\includegraphics[width=1.8\columnwidth]{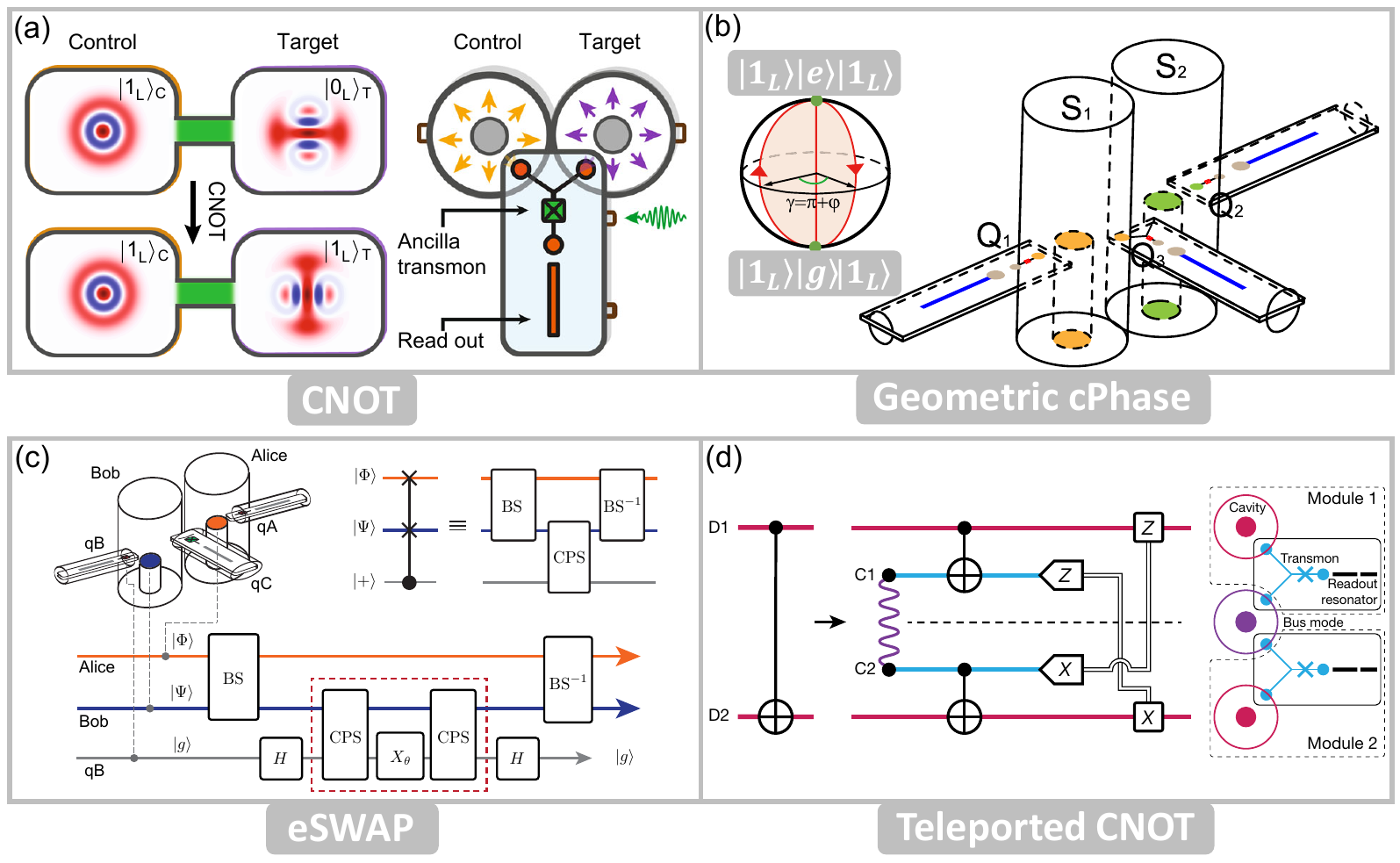}
\caption{\textbf{Unitary gates on two logical qubits.} {(a)} Controlled-not (CNOT) gate between two logical qubits with the target one is binomially encoded. Adapted from Ref.~\cite{Rosenblum2018}. {(b)} Geometrically controlled phase gate on two binomially encoded logical qubits. Adapted from Ref.~\cite{XuPRL2020PhotonicQubits}. {(c)} Exponential-SWAP
gate on two bosonic qubits. Adapted from Ref.~\cite{GaoNature2019}.  {(d)} Teleported CNOT gate between two binomially encoded logical qubits. Adapted from Ref.~\cite{Chou2018}.}
\label{fig:LogicalGates}
\end{figure*}

\subsection{Binomial codes}
\label{sec:BinomialCodes}
Binomial codes are based on superpositions of truncated Fock states weighted with binomial coefficients~\cite{Michael2016}. These codes are designed to exactly correct errors that are polynomial up to a specific order in photon loss error $\hat{a}$, photon gain error $\hat{a}^{\dagger}$, and photon dephasing error $\hat{n}$, i.e., the error set is:
\begin{equation}
\varepsilon=\{\hat{I},\hat{a},\hat{a}^{2},...,\hat{a}^{L},\hat{a}^{\dagger},(\hat{a}^{\dagger})^{2},..,(\hat{a}^{\dagger})^{G},\hat{n},\hat{n}^{2},...,\hat{n}^{D}\}.
\end{equation}
The code basis states are:
\begin{equation}
\begin{aligned}\ket{0_{L}}=\frac{1}{\sqrt{2^{N}}}\sum_{p,even}^{[0,N+1]}\sqrt{C_{N+1}^{p}}\ket{p(S+1)},\\
\ket{1_{L}}=\frac{1}{\sqrt{2^{N}}}\sum_{p,odd}^{[0,N+1]}\sqrt{C_{N+1}^{p}}\ket{p(S+1)},
\end{aligned}
\label{eq:binomialcodes}
\end{equation}
where $C_{N+1}^{p}$ is the binomial coefficient, the spacing is $S=L+G$, $N=\mathrm{max}\{L,G,2D\}$, and $p$ is from 0 to $N+1$ with the maximum Fock number being $(N+1)\times(S+1)$.

It can be clearly seen that the two basis states contain completely different series of Fock states, therefore are exactly orthogonal to each other. It is also easy to check that both states have exactly the same average photon numbers. As a result, the QEC conditions Eqs.~\ref{eq:QEClogicalstateindis}
and \ref{eq:QEClogicalstateOrthognonal} are strictly satisfied. The spacing of the occupied Fock states is $S+1$, therefore, the errors can be uniquely distinguished by measuring photon number modulo $S+1$, i.e., the error syndrome is the generalized parity. Because the basis states in the code space and all error spaces are orthogonal, the binomial codes have the advantage of having explicit unitary operations for repumping energy into the mode when compared to the cat codes. Besides, since the occupied Fock states are truncated, unitary operations on the binomial codes might also be more convenient.



Figure~\ref{fig:QEC_CatBinomialGKP}b shows the lowest-order binomial code:
\begin{equation}
\begin{aligned}\ket{0_{L}} & =(\ket{0}+\ket{4})/\sqrt{2},\\
\ket{1_{L}} & =\ket{2},
\end{aligned}
\label{eq:LowestBincodes}
\end{equation}
This code can protect against single-photon-loss error with $\varepsilon=\{\hat{I},\hat{a}\}$. The average photon number is two, smaller than that of the typical cat code. When an error occurs, the corresponding error space is:
\begin{equation}
\begin{aligned}\ket{0_{E}}=\ket{3},\\
\ket{1_{E}}=\ket{1}.
\end{aligned}
\label{eq:LowestBincodesError}
\end{equation}
Clearly, these two basis states have different average photon numbers, not satisfying the QEC condition Eq.~\ref{eq:QEClogicalstateindis} anymore. A unitary recovery operation has to be applied immediately
to correct the error, otherwise, quantum information will be corrupted. This property is different from the previously discussed cat codes.

A comparison between the lowest-order binomial code with the qubit-based four-qubit code~\cite{Leung1997} can better shed light on the efficiency and advantage of the bosonic codes. The four-qubit code can correct single amplitude damping errors, $\varepsilon=\{\hat{I},\hat{\sigma}_{1}^{-},\hat{\sigma}_{2}^{-},\hat{\sigma}_{3}^{-},\hat{\sigma}_{4}^{-}\}$,
with the basis states:
\begin{equation}
\begin{aligned}\ket{0_{L}}=\frac{1}{\sqrt{2}}(\ket{0000}+\ket{1111}),\\
\ket{1_{L}}=\frac{1}{\sqrt{2}}(\ket{1100}+\ket{0011}).
\end{aligned}
\label{eq:4qubitcode}
\end{equation}
This encoding utilizes $2^{4}=16$ dimensional expanded Hilbert space. In order to uniquely distinguish the five errors in the error set, three error syndromes are required. In marked contrast, although both the lowest-order binomial code and the four-qubit code have the same average excitation of two, the binomial code occupies only the lowest five levels of the oscillator's Hilbert space (five dimensions) and needs only one error syndrome for error detection. Therefore, the bosonic codes are indeed hardware-efficient and can greatly save physical resources.

According to Eq.~\ref{eq:binomialcodes}, in order to correct more errors, for example, $\varepsilon=\{\hat{I},\hat{a},\hat{a}^{2},\hat{n}\}$, one has to use a higher-order binomial code with a larger Fock state dimension for the encoding:
\begin{equation}
\begin{aligned}\ket{0_{L}} & =\frac{\ket{0}+\sqrt{3}\ket{6}}{2},\\
\ket{1_{L}} & =\frac{\sqrt{3}\ket{3}+\ket{9}}{2}.
\end{aligned}
\label{eq:SecondBincodesError}
\end{equation}
The spacing of the occupied Fock states is three, i.e. the logical states are conserved under the generalized parity operator $\hat{\Pi}=e^{i\frac{2}{3}\pi\hat{a}^{\dagger}\hat{a}}$. \red{A photon gain error and two-photon-loss errors have the same change in the photon number modulo 3.} As a result, the above code can also correct errors in the set $\varepsilon=\{\hat{I},\hat{a},\hat{a}^{\dagger},\hat{n}\}$. \red{For large average photon number in the codewords, the binomial and cat codes asymptotically approach each other since both photon number distributions approach a normal distribution~\cite{AlbertPRA2018}.}

It is worth noting that even when no error is detected (no photon loss error) the binomial codes still suffer the non-unitary backaction associated with no-error evolution $e^{-(\kappa/2)\hat{a}^{\dagger}\hat{a}t}$ (see Eq.~\ref{eq:ExactPhotonLossError} for the exact expression
of photon loss errors). This is a nontrivial distortion of the code states and must be corrected. \red{A two-mode version of the codes with the same spacing and the same total excitation number but distributed
in different modes can mitigate this problem~\cite{Michael2016,Chuang1997}.}

Experimentally, the lowest-order binomial code (Eq.~\ref{eq:LowestBincodes}) has been demonstrated with repetitive QEC based on realtime feedback control~\cite{Hu2019}, as shown in Fig.~\ref{fig:QEC_CatBinomialGKP}e. The QEC protected quantum information has a lifetime nearly beating the break-even point. A high-fidelity universal gate set operation on the logical qubit has also been demonstrated. Towards universal quantum computation based on binomial codes, a cPhase gate between two binomial logical qubits has been realized through a geometric method~\cite{XuPRL2020PhotonicQubits} (Fig.~\ref{fig:LogicalGates}b). A teleported CNOT gate between two binomial logical qubits and a CNOT
gate with the target being a binomial logical qubit have also been
realized~\cite{Chou2018,Rosenblum2018}, as illustrated in Figs.~\ref{fig:LogicalGates}a and \ref{fig:LogicalGates}c, respectively.

\subsection{GKP codes}
\label{sec:GKPcodes}

The GKP codes were first proposed by Gottesman, Kitaev, and Preskill
in 2001~\cite{GottesmanPRA2001}. The general GKP codes can protect
a state of a $d$-dimensional quantum system (a qudit) encoded
in a harmonic oscillator against most physical noise precesses. For a typical
two-level logical qubit, the GKP code is defined as coherent superpositions
of infinitely squeezed states or the eigenstates of the position operator
$\hat{q}$ with a spacing of $2\sqrt{\pi}$:
\begin{equation}
\begin{aligned}\ket{0_{L}} & \propto\sum_{s=-\infty}^{\infty}\ket{q=2s\sqrt{\pi}},\\
\ket{1_{L}} & \propto\sum_{s=-\infty}^{\infty}\ket{q=(2s+1)\sqrt{\pi}}.
\end{aligned}
\label{eq:GKPcodeword}
\end{equation}
These two basis states are shifted by $\sqrt{\pi}$ relative to each
other and their corresponding Wigner functions in the $q-p$ phase space
are square grid patterns.

It is known that the GKP codes belong to the class of stabilizer codes.
The grid states of Eq.~\ref{eq:GKPcodeword} are in fact stabilized
by two mutually commuting stabilizers:
\begin{equation}
\begin{aligned}
\hat{S}_{q} & =\hat{D}(i\sqrt{2\pi})=e^{i2\sqrt{\pi}\hat{q}},\\
\hat{S}_{p} & =\hat{D}(\sqrt{2\pi})=e^{-i2\sqrt{\pi}\hat{p}},
\end{aligned}
\label{eq:GKP_stabilizers}
\end{equation}
where $\hat{D}(\alpha)=e^{\alpha\hat{a}^{\dagger}-\alpha^{*}\hat{a}}$
is the displacement operator and $\hat{p}$ is the momentum operator. Consequently, the GKP code space is the simultaneous eigenspace of the above two stabilizers. The corresponding Pauli operators are simple displacements of half the grid spacing:
\begin{equation}
\begin{aligned}X & =\hat{D}(\sqrt{\pi/2})=e^{-i\sqrt{\pi}\hat{p}},\\
Z & =\hat{D}(i\sqrt{\pi/2})=e^{i\sqrt{\pi}\hat{q}},\\
Y & =\hat{D}(\sqrt{\pi/2}+i\sqrt{\pi/2}),
\end{aligned}
\label{eq:GKP_stabilizers}
\end{equation}
satisfying the Pauli relations. 

In the Fock state representation, the ideal GKP codewords contain infinite number of photons and correspond to Wigner functions that extend to infinity in phase space. Therefore, the ideal GKP codes are not physical. The realistic GKP states have finite photon energy and are approximate by replacing the delta functions with
finitely squeezed Gaussian state and the uniform superposition profile
with a Gaussian envelope centered around $q=0$. The approximate GKP
codewords therefore become~\cite{GottesmanPRA2001,Terhal_2016}:
\begin{equation}
\begin{aligned}\ket{0_{L}}_{\mathrm{approx}} & \propto\sum_{s=-\infty}^{\infty}e^{-2\pi\tilde{\Delta}^{2}s^{2}}\hat{D}(s\sqrt{2\pi})\ket{\psi_{0}},\\
\ket{1_{L}}_{\mathrm{approx}} & \propto\sum_{s=-\infty}^{\infty}e^{-\pi\tilde{\Delta}^{2}(2s+1)^{2}/2}\hat{D}(s\sqrt{2\pi})\hat{D}(\sqrt{\pi/2})\ket{\psi_{0}},
\end{aligned}
\end{equation}
where $1/\tilde{\Delta}$ is the width of the Gaussian envelope and
$\ket{\psi_{0}}=\int\frac{dq}{(\pi\Delta^{2})^{1/4}}e^{-q^{2}/(2\Delta^{2})}\ket{q}$
is the squeezed vacuum state with $\Delta$ being the squeezing parameter.
The corresponding Wigner functions in the $q-p$ phase space are shown
in Fig.~\ref{fig:QEC_CatBinomialGKP}c.

The square GKP codewords have asymmetric error-resistance property because of the asymmetric nature of the three Pauli operators defined in Eq.~\ref{eq:GKP_stabilizers}. To get a symmetric protection against errors in all three directions, the lattice of the square code can be transformed into a hexagonal code. The hexagonal GKP code may be the ultimate optimal code, because starting from a random initial code, numerical optimization for both a photon loss channel and a Gaussian thermal loss channel always converges to the hexagonal GKP code~\cite{Noh_2019}. In addition, compared to other bosonic QEC codes including cat codes, binomial codes, and numerically optimized codes, the GKP codes show the best performance for most values of the photon loss rate~\cite{AlbertPRA2018}. \red{However, for the same average photon number, the GKP codes have a larger bandwidth of photon number distribution or a larger occupied Hilbert space, and thus suffer more distortion from Kerr effect. This poses an experimental challenge to high-fidelity recovery and demands further optimization of the codes including the Kerr effect~\cite{Li2019}.}


The GKP codes are designed to correct small shift errors as long as $|\delta q|<\sqrt{\pi}/2$ and $|\delta p|<\sqrt{\pi}/2$. Measurements of the stabilizers (the error syndromes) unambiguously reveal the underlying
errors, which can be corrected by shifting back with the minimal amount of displacement. A variety of local errors, such as photon loss, thermal noise, photon dephasing, and even spurious nonlinearities induced by the coupled ancilla qubit, lead to a continuous evolution of the states in phase space, and hence result in only local effects in the phase space. Therefore, as long as the stabilizers are measured frequently enough, the noise-induced shifts will be small and thus can be detected and corrected. In fact, it is also shown that the local errors can be expanded into small shift errors in $p$ and $q$ when the number of photons in the GKP codewords are small~\cite{Terhal_2016}.

Another advantage of the GKP codes is that the Clifford gates only require Gaussian operations on the photonic state, which is usually easy to perform in the experiment. Under these gate operations, small deviations of $q$ and $p$ remain small, which means the locality of the errors is preserved. In this sense, these gates are fault tolerant.

However, the non-Clifford gates are much harder than the Clifford gates, demanding non-Gaussian operations or resources. One method is to prepare a magic state such as the eigenstate of the Hadamard gate, and use it as an ancilla. Then only Clifford gates and homodyne measurement are required to perform the non-Clifford $\hat{T}$ gate~\cite{GottesmanPRA2001}. The preparation of the codewords is also challenging and requires non-Gaussian operations.

\red{It is hard to scale up the GKP codes only by increasing the squeezing rate in a practical physical system. Besides, the GKP codes are not designed to protect against rare and large errors. Therefore, the GKP codes are usually considered to concatenate with other stabilizer codes for a second layer of protection~\cite{yamasaki2020polylogoverhead,Terhal_2020}, for example, the surface codes. The QEC process for these stabilizer codes only requires Clifford gates and homodyne measurement. Several theoretical works~\cite{Fukui_2018,wang2019quantum,Noh_2020,Terhal_2020} calculate
the required squeezing level, about 10-20 dB, to reach the fault-tolerant threshold of the surface-GKP code based on different assumptions on the error source.}

Although the GKP codes were proposed early, experimental demonstrations of the GKP codes have been realized only very recently. Encoding, logical readout, and full control of a GKP qubit have been demonstrated in the motional mode of a single trapped ion~\cite{Fl_hmann_2019}. QECs of both square and hexagonal GKP codes have been demonstrated in a superconducting microwave cavity~\cite{campagneibarcq2020}, where the GKP code states can be deterministically generated from a vacuum state based on repeated stabilizer measurements \red{and QEC protocol} facilitated with feedback technique~\cite{Terhal_2016}. Continuous QEC on the GKP qubit has shown the extension of the coherence of the logical qubit, demonstrating the capability of suppressing all logical errors.

\begin{figure*}
\centering{}\includegraphics[width=1.8\columnwidth]{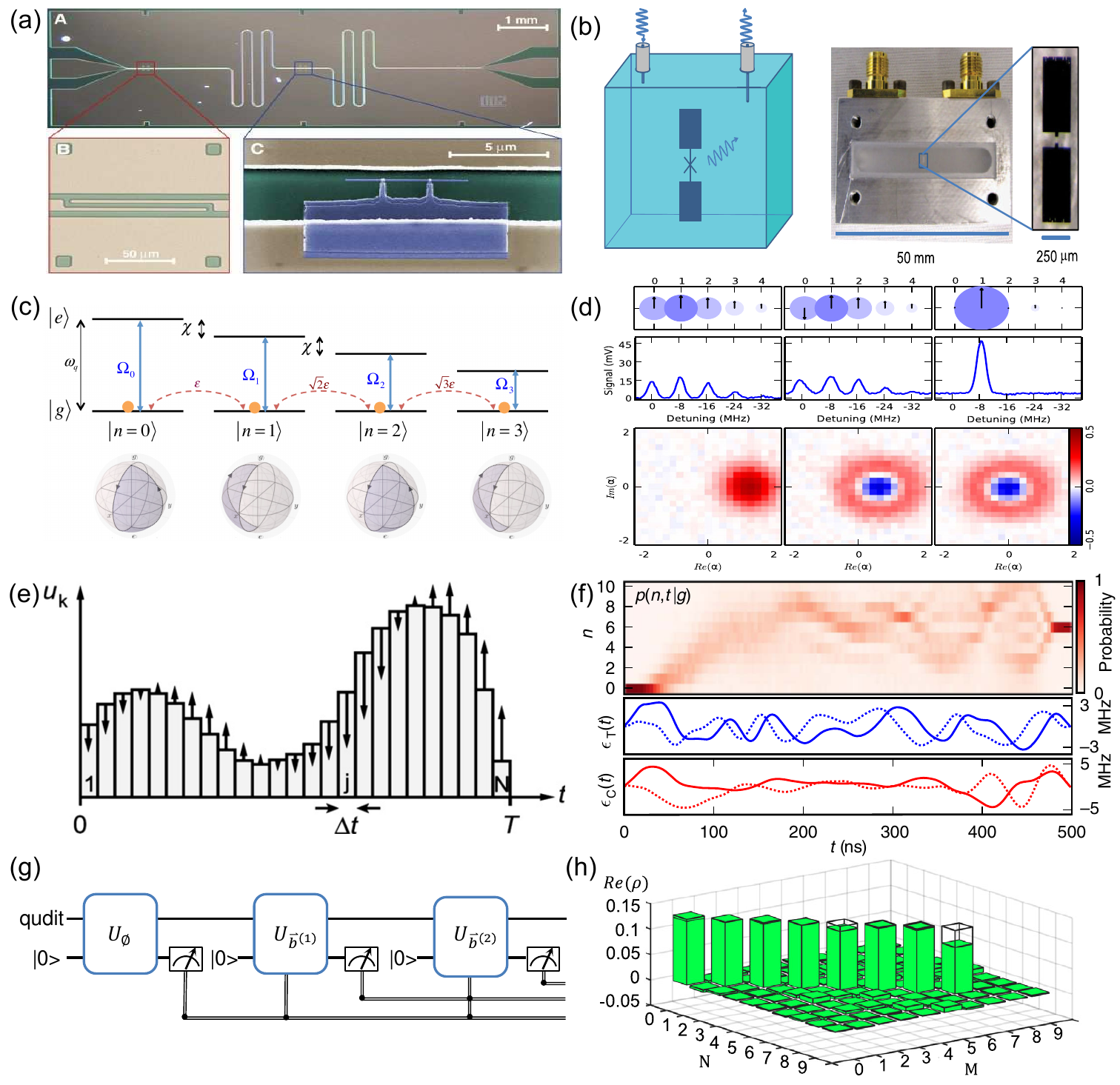}
\caption{\textbf{Universal control of individual bosonic mode.} {(a)} First
demonstration of circuit QED architecture. Adapted from Ref.~\cite{Wallraff2004} {(b)} First demonstration of a 3D circuit QED architecture. Adapted from Ref.~\cite{Paik2011}. {(c)} Principle of the SNAP gate. Adapted from Ref.~\cite{Krastanov2015}. {(d)}
Experimental results of a SNAP gate. Adapted from Ref.~\cite{HeeresPRL2015}. {(e)} Schematic representation of a control amplitude consisting of $N$ steps in the GRAPE method. Adapted from Ref.~\cite{Khaneja2005}. {(f)} Experimental results of the Fock state population evolution and the corresponding GRAPE pulses. Adapted from Ref.~\cite{Heeres2017}. {(g)} Quantum circuit for arbitrary
channel construction with adaptive control. Adapted from Ref.~\cite{ShenPRB2017}. {(h)} Density matrix of maximally-mixed state $\sum_{k=0}^{7}|k\rangle\langle k|$ generated based on the quantum circuit in (g). Adapted from Ref.~\cite{Wang2019DQC1}.}
\label{fig:UniversalControl}
\end{figure*}

\section{Universal quantum control of bosonic codes}
\label{sec:UniversalControl}
In the realization of bosonic codes, including encoding, decoding, universal gate set, and error detection and correction operations, universal quantum control of a bosonic mode is crucial. Such a goal is important for not only QEC, but
also the understanding and controlling of quantum systems. Inspired by the cavity QED experiments~\cite{Haroche,Haroche2020}, the universal quantum control could be achieved in a so-called spin-oscillator model by introducing a two-level system to couple to the bosonic mode. Circuit QED has been extensively studied in the past decades and has become one of the most promising platforms for quantum computing~\cite{Blais2004,Wallraff2004,You2011,Devoret2013,Gu2017,BlaisNP2020,Blais2020cQED}.
Tremendous progress on the control of a bosonic mode has been made in this architecture since its first development in Ref.~\cite{Wallraff2004}
(Fig.~\ref{fig:UniversalControl}a). Here, the results are summarized
in two parts for unitary quantum control of closed systems and
quantum channels for open systems, respectively.


The Hamiltonian of a typical circuit QED system consisting of a
transmon qubit and a cavity mode in the largely detuned regime can
be described by~\cite{Devoret2013,Kirchmair2013,Vlastakis2013,SunNature2014,Blais2020cQED}
\begin{equation}
\hat{H}_{0}/\hbar=\omega_{c}\hat{a}^{\dagger}\hat{a}+\omega_{q}\left|e\right\rangle \left\langle e\right|-\chi \hat{a}^{\dagger}\hat{a}\left|e\right\rangle \left\langle e\right|-\frac{K}{2}\hat{a}^{\dagger2}\hat{a}^{2}.\label{eq:Hamiltonian}
\end{equation}
Here, $\omega_{c}$ and $\omega_{q}$ are the cavity and qubit frequencies,
respectively, $\hat{a}^{\dagger}$ ($\hat{a}$) is the creation (annihilation)
operator for the bosonic mode, $\ket{e}$ ($\ket{g}$) is the
excited (ground) state of the transmon qubit, and $\chi$ and $K$ are
the dispersive coupling and Kerr coefficient originating from the
qubit, respectively. The two-level qubit serving as an ancilla provides
the necessary non-linearity for the universal control of not only the
bosonic mode but also the whole combined system.

The strong dispersive coupling allows the resolving of Fock states and thus the implementation of photon-number-selective operations. Universal control of the bosonic mode can be achieved by using selective number-dependent arbitrary phase gates (SNAP)~\cite{Krastanov2015} in combination with displacement operations. The SNAP gate reads:
\begin{equation}
\hat{S}(\vec{\theta})=\sum_{n=0}^{\infty}e^{i\theta_{n}}|n\rangle\langle n|.
\end{equation}
Here $\vec{\theta}=\{\theta_{n}\}_{n=0}^{\infty}$ is a list of phases
and $\left|n\right\rangle $ is the $n$-photon Fock state. Each phase
gate is generated geometrically as shown in Fig.~\ref{fig:UniversalControl}c.
These geometric phase gates preserve photon numbers, while displacement
operations induce the hopping between adjacent Fock states. The combination
of both operations gives a universal control of the cavity. For example,
a unitary $\hat{U}_{n}$ that transfers the population between $\ket{n}$ and
$\ket{n+1}$ is given by
\begin{equation}
\hat{U}_{n}=\hat{D}(\alpha_{1})\hat{R}_{n}(\pi)\hat{D}(\alpha_{2})\hat{R}_{n}(\pi)\hat{D}(\alpha_{3}),
\end{equation}
where $\hat{R}_{n}(\pi)=-\sum_{n'=0}^{n}|n'\rangle\langle n'|+\sum_{n'=n+1}^{\infty}|n'\rangle\langle n'|$
is the SNAP gate, and $\alpha_{1},\alpha_{2},\alpha_{3}$ in the displacement
operators can be optimized to maximize the fidelity $|\langle n+1|\hat{U}_{n}|n\rangle|$. A one-photon Fock state $\ket{1}$ has been experimentally generated with this method~\cite{HeeresPRL2015} (Fig.~\ref{fig:UniversalControl}d). A similar idea based on the photon-number-selective detection for arbitrary state preparation has also been proposed and demonstrated~\cite{Wang2017}: A cavity initially prepared in a coherent state can be projected into an arbitrary superposition of Fock states with high fidelities but in a probabilistic manner by post-selecting the measurement outcomes of the ancilla.

The SNAP gate method requires to control the qubit and the oscillator separately with a series of sequential SNAP gates and displacement operations. To overcome the drawback of the relatively long gate time, a more efficient method is proposed
recently which involves a hierarchical insertion strategy and gradient-descent
technique for parameter optimization and shows remarkable improvement~\cite{Fosel2020}.



Another efficient approach for realizing universal control is the
optimal control technique, which explores the full control parameter
space to optimize the control pulses and has been widely used in experiment.
Figure~\ref{fig:UniversalControl}e shows the schematic of the gradient
ascent pulse engineering (GRAPE) method~\cite{Khaneja2005,DeFouquieres2011}
to optimize control pulses of a target unitary. The evolution time
is discretized into small steps and the initial controls could be
completely random. The performance function is based on the forward
and backward propagations of the initial and target density operators,
respectively. At each step, the gradients of the performance function
with respect to the controls are calculated and a gradient ascent
procedure is followed by updating the control parameters to improve the performance function. Because of its universality and simplicity, \red{the GRAPE method
has been extensively applied in the encoding of QEC codes
and the unitary gates on the logical qubits~\cite{Heeres2017,Hu2019,Gertler2020}
(see Fig.~\ref{fig:UniversalControl}f as an example).}

The spin-oscillator model could be extended to multiple modes
by introducing dispersive couplings between the central ancilla qubit
and more modes, as well as the direct cross-Kerr nonlinearities between
modes due to the qubit. As an example, the simplest two-cavity SNAP
gate is demonstrated in Ref.~\cite{XuPRL2020PhotonicQubits}, where
a single-photon Bell state between two cavities is deterministically
generated by inducing a $\pi$-phase shift on the $\left|0\right\rangle \otimes\left|0\right\rangle $ state of the two cavities. Figure~\ref{fig:LogicalGates} summarizes the recently demonstrated two-cavity gates for bosonic codes with various approaches~\cite{Rosenblum2018,Chou2018,GaoNature2019,XuPRL2020PhotonicQubits}. Therefore, the universal gate set on bosonic codes is available. However, although both SNAP and GRAPE approaches could be generalized to gates between logical qubits based on arbitrary bosonic codes, the significant increase of the system Hilbert space imposes great challenges in numerical optimization of the control pulse sequences and also difficulties in experimental realization.

The above two approaches for universal control are restricted to closed quantum systems. However, practical quantum systems are open due to their inevitable coupling to the environment, and their dynamics are described by completely positive and trace preserving quantum channels~\cite{Nielsen2000}. So besides the universal control of a closed quantum system, the realization of arbitrary quantum
channels helps to understand practical quantum systems and complete
our capability in quantum controls. For instance, the QEC process is a quantum channel that purifies the quantum state of a system by removing the entanglement between the system and the environment. A universal approach for quantum channel simulations of a bosonic mode has been proposed~\cite{Lloyd2001PRA,ShenPRB2017}, holding the hardware-efficiency advantage for bosonic codes and being promising for a wide range of applications of bosonic modes, such as system initialization, generalized quantum measurements, open quantum system simulation, and quantum metrology. Figure~\ref{fig:UniversalControl}g illustrates the kernel idea of the approach: by repetitively using and resetting the ancilla qubit and also using the output of the ancilla measurement for feedforward control of the bosonic mode, arbitrary quantum channels can be implemented.

Preliminary experimental studies on arbitrary quantum channel simulations of a photonic qubit in a superconducting circuit, which is encoded in the first two levels of an oscillator, have been demonstrated~\cite{Hu2018channel}. In this experiment, the arbitrary single-qubit channel simulations require a fast real-time feedback control system for adaptive operations conditional on the specific measurement results. A quantum channel for maximally-mixed state
preparation is demonstrated in Ref.~\cite{Wang2019DQC1}, with the
results shown in Fig.~\ref{fig:UniversalControl}h. In a different experiment, a specific quantum channel, i.e. QEC operation on a binomial code, is implemented without the feedback electronic circuit. This autonomous QEC (AQEC) does not need to extract error detection outcomes~\cite{Ma_2020}. Instead, a unitary transition $\hat{U}$ is implemented to transfer the error entropy associated with the logical state to the ancilla and in the meantime the logical state in the error space $\ket{\psi_{\mathrm{E}}}$ is converted back to the correct one $\ket{\psi_{\mathrm{L}}}$ in the code space as:
\begin{equation}
\begin{aligned}
\hat{U}|\psi_{\mathrm{E}}\rangle|g\rangle=|\psi_{\mathrm{L}}\rangle|e\rangle,\\
\hat{U}|\psi_{\mathrm{L}}\rangle|g\rangle=|\psi_{\mathrm{L}}\rangle|g\rangle.
\end{aligned}
\label{eq:AQEC}
\end{equation}
Therefore, the correlation between the logical state and the environment
(which induces errors) is erased. Since the real-time feedback control system
is not required, the potential electronic latency is avoided. A
separate experiment realizes AQEC of single-photon-loss errors on a
so-called truncated 4-component cat code~\cite{Gertler2020}. The
unitary transition is realized through two combs of continuous and
selective microwave drives with no which-path information leaking
into the environment, while the ancilla reset is through a dissipative
process. \red{Achieving the full control of an open quantum system
is necessary for the bosonic codes, and more experimental efforts are
required in this direction for more advanced quantum control.} For
example, besides the standard error correction in an autonomous manner, fault tolerance to ancilla errors is also possible by carefully designing the control (see Sec.~\ref{sec:FaultTolerance} for more discussions).

Lastly, it is also worth noting that other than the spin-oscillator
model widely studied in the superconducting quantum circuit, \red{the Pockel and Kerr nonlinearities of harmonic oscillators, which originate from the intrinsic bulky material nonlinearity,}
also hold the potential for universal control of the bosonic modes. As
widely studied in the continuous variable quantum information,
these bulky nonlinearities with modest interaction strength could simulate
arbitrary Hamiltonian via a Trotterization approach~\cite{Weedbrook2012}. These
nonlinearities are more suitable for continuous variable encodings because there is
no requirement for the approximation of truncated Fock
space. The possibility of using the Pockel nonlinearity
in universal quantum computation is confirmed in recent theoretical
studies~\cite{Niu2018}, which provides an alternative route to applications of
bosonic codes.




\begin{figure*}
\centering{}\includegraphics[width=1.8\columnwidth]{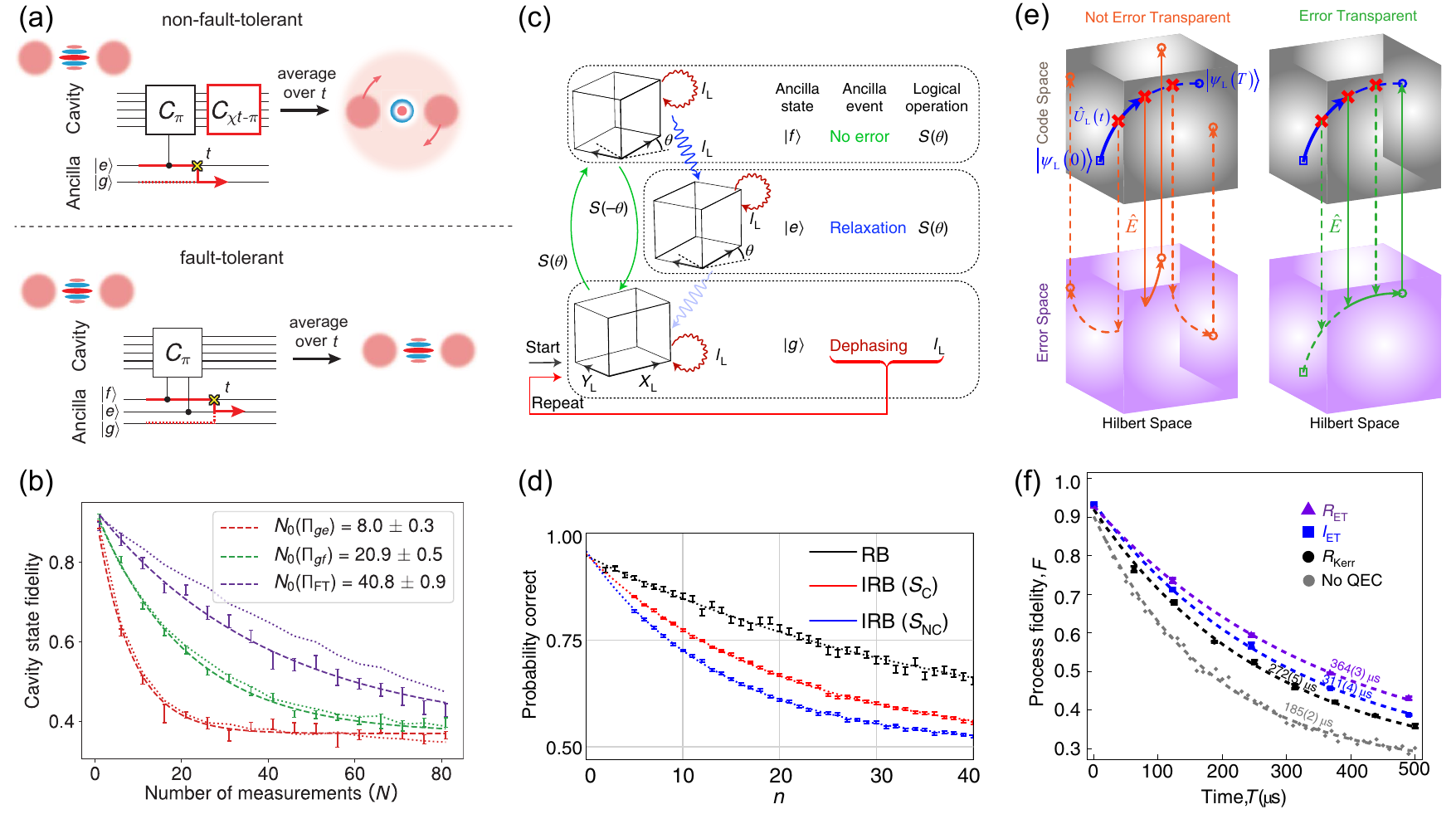}
\caption{\textbf{Fault tolerant operations of bosonic codes.} (a) Schematic
of a fault-tolerant quantum error detection, where an auxiliary energy level $\ket{f}$ of the ancilla is employed. (b) Experimental results of the fault-tolerant error detection show a suppression of the ancilla errors by
a factor of five. (a-b) are adapted from Ref.~\cite{Rosenblum_2018}.
(c) The principle of path-independent phase gate. Path independence
requires the ancilla is manipulated in such a way that its populations
are independent of the state of the encoded system. (d) Benchmarking
results of the logical gate demonstrate a significant improvement
due to the path-independent design. (c-d) are adapted from Ref.~\cite{reinhold2020errorcorrected}. (e) Concept of the
error-transparent gate. The tracks of quantum evolution in both the code and the error spaces are deterministic and identical irrespective of the time when the error occurs. (f) Experimental process fidelity as a function of time with repetitive and interleaved error-transparent gates and autonomous QEC on the logical qubit demonstrates an improved performance. (e-f) are adapted from Ref.~\cite{Ma_2020}.}
\label{fig:FaultTolerantOperations}
\end{figure*}


\section{Applications of bosonic codes}
\label{sec:Applications}
As all ingredients of bosonic codes are available in superconducting
quantum circuits, their direct applications in storing and transferring
quantum information, i.e. in realizing quantum computation and quantum
communication, are foreseeable. From another perspective, a single
bosonic mode supports an infinitely large Hilbert space, and thus provides
a unique platform for exploring quantum advantages in quantum simulation
and metrology. However, in practical near-term noisy intermediate-scale quantum (NISQ)~\cite{Preskill2018} platforms, we should not be restricted to
the standard QEC codes that satisfy the QEC condition (Eq.~\ref{eq:QECconditions}). In certain tasks, the bosonic codes are beneficial to reduce the effects of noise
and system imperfections on the estimation of certain outputs by
QEC or approximate QEC. Although great advantages are
promised by bosonic codes, only preliminary experimental and theoretical
results are reported. The potentials of bosonic codes are awaiting
systematic investigations with many techniques and theoretical problems to be solved. Here, we just summarize the recent exciting progress and proof-of-principle demonstrations, and point out the opportunities in future studies.

\subsection{Fault-tolerant quantum computation}
\label{sec:FaultTolerance}

QECs are developed to protect merely stored quantum information from
the leading orders of errors. However, for a general purpose of quantum
information processing, the errors occurring during the dynamical evolution
might not be correctable by directly applying QEC after the gate.
Therefore, a fault-tolerant universal quantum computer is also required
to protect the dynamics of quantum information during each step of
computing, which should be carefully designed to keep errors from
propagating and accumulating such that each encoded logical qubit
can still be well protected by QECs. In another words, state preparations,
error detections, gate operations, and measurements are all needed
to be fault tolerant. For qubit-based systems, surface code architecture and code-concatenation approaches are proposed for achieving the ultimate fault tolerance and \red{a clear threshold of the error rate is provided for reliable and scalable quantum computation}~\cite{Fowler2012,Jochym-OConnor2014}. However, these schemes are extremely challenging for experimental realization because they require huge physical sources that are not available currently. In contrast, benefiting from the hardware-efficiency property, the encoding, decoding, error corrections, and universal logical gate set on encoded logical qubits have been achieved with the bosonic codes. The experimental explorations of fault-tolerant operations on the bosonic codes are already in progress, and Fig.~\ref{fig:FaultTolerantOperations} summarizes some of the results. Note that the early attempts towards the fault-tolerant quantum computation are the demonstrations of the literal meaning of fault tolerance, i.e. the capability of correcting certain physical errors occurring during the gate operation, \red{instead of achieving the fault-tolerant threshold.}



Because the ancilla plays a significant role in realizing the operations
on bosonic codes, the damping and dephasing errors of the ancilla
might induce significant errors on the bosonic codes. For example,
error detection on cat codes and binomial codes is a non-Gaussian
operation and thus requires an ancilla (unlike that for GKP codes). Qubit damping error $\hat{\sigma}^{-}$ in the error-detection circuit will
propagate to the encoded information by causing random phase-shifts
which cannot be corrected, as illustrated in Fig.~\ref{fig:FaultTolerantOperations}a. Fault-tolerant error detection hence demands the prevention of the ancilla error from propagating to and corrupting the encoded system. By introducing redundant energy levels of the ancilla, a fault-tolerant error detection scheme against the ancilla damping error is demonstrated~\cite{Rosenblum_2018}. This scheme is similar to use a QEC-protected ancilla system, and the experiment demonstrates a suppression of the ancilla errors by a factor of five.


An alternative way is to use an ancilla qubit with biased-noise~\cite{PuriPRX2019}.
As the operator $\hat{\sigma}_{z}$ commutes with the interaction Hamiltonian ($\chi \hat{a}^{\dagger}\hat{a}\hat{\sigma}_z/2$), an ancilla with only $\hat{\sigma}_{z}$ error will not damage the encoded system, but only influence the detection result. Cat qubits under continuous parametric drive are one candidate of realizing such biased-noise qubits. The phase-flip ($\hat{\sigma}_{z}$) rate is only linearly enhanced but the bit-flip ($\hat{\sigma}_{x}$) rate is exponentially suppressed with the size of the cat qubit. Using such a stabilized cat qubit as the ancilla,
without intrinsic errors that do not commute with the interaction, the simulation in Ref.~\cite{PuriPRX2019} shows the measurement backaction on the encoded system can indeed be suppressed.



\begin{figure*}
\centering{}\includegraphics[width=1.8\columnwidth]{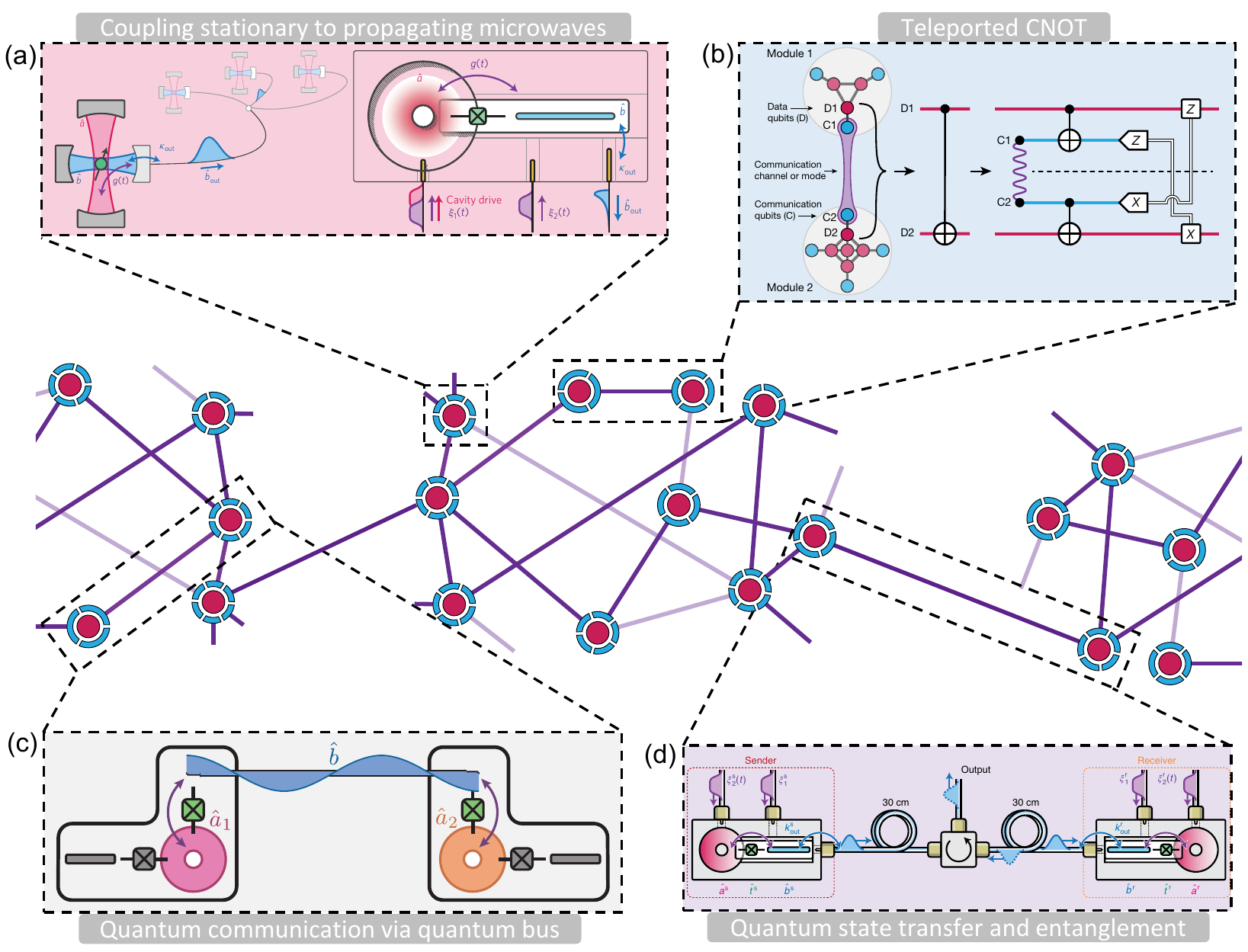}
\caption{\textbf{Quantum communication via a quantum network.} Each node or module
represents a small quantum processor consisting of data qubits and
communication qubits. {(a)} Controlled release of photonic states
from one of the module. Adapted from Ref.~\cite{Pfaff2017}. {(b)} Teleported CNOT circuit between two modules. Adapted from Ref.~\cite{Chou2018}. {(c)} Quantum communication between two modules via a quantum bus. Adapted from Ref.~\cite{burkhart2020errordetected}.
{(d)} On-demand quantum state transfer and entanglement. Adapted from Ref.~\cite{Axline2017}.}
\label{fig:QuantumCommunications}
\end{figure*}

To perform gate operations on the bosonic codes, the ancilla system
is also necessary. To prevent error propagation from the ancilla system
to the encoded system, a theoretical work analyzes the conditions
on the interaction Hamiltonian and defines the concept of ``path independence"~\cite{ma2019pathindependent}. When the ancilla system starts from $\ket{i}$ and ends in $\ket{r}$, the $n$-th order path-independent gate requires that the encoded system evolves under a deterministic unitary even when the ancilla
system suffers errors up to the $n$-th order. A subset of final ancilla
states indicate the successful implementation of the desired gate, while
other states herald a failure of the operation, but the encoded system
is not corrupted in this process. The additional drives and the measurement
to distinguish additional levels of the ancilla, however, might introduce more
error sources. A path-independent phase gate with the SNAP technique
is demonstrated in the experiment~\cite{reinhold2020errorcorrected},
where the fidelity of the SNAP gate on a three-level ancilla qubit
is significantly improved by the specific path-independent design,
as shown in Fig.~\ref{fig:FaultTolerantOperations}b.




Besides the tolerance of ancilla errors during the desired gate operations,
the errors occurring in the encoded system should also be considered
and prevented from propagation. To ensure photon loss error will not propagate under arbitrary unitary evolutions, the concept of ``error-transparent" gate is introduced~\cite{vy2013error,kapit2018error}. The basic idea is the following. Ideally, a quantum state should evolve unitarily in the code space under the gate Hamiltonian with a finite gate time. If an error happens during the gate, the evolution will jump to the error space, while the subsequent evolution in the error space is identical to that in the code space up to a global phase. So this
error during the gate operation is tolerable by QEC at the end of the evolution, and the gate can still be implemented successfully. A recent experiment has demonstrated error-transparent phase gates on the lowest-order binomial code~\cite{Ma_2020}. States in both the code and the error spaces are preserved and the lifetime of the QEC-protected logical state has better performance under error-transparent gate operation, as shown in Fig.~\ref{fig:FaultTolerantOperations}c. In Ref.~\cite{Ma_2020}, the authors also show that the error-transparent gates could be generalized to a universal gate set. Further extension of this approach could be combined with the AQEC technique. To prevent the ancilla error propagation in this process, one method is to design the AQEC Hamiltonian as follows:
\begin{equation}
\begin{aligned}
\hat{H}=\sum_{ij}|L_{i}\rangle|j\rangle\langle0|\langle E_{ij}|+h.c.,
\end{aligned}
\end{equation}
where $\ket{E_{ij}}$ and $\ket{L_{i}}$ are the $i$-th logical basis state
in the $j$-th error space and the code space respectively, and $\ket{0}$
and $\ket{j}$ is the ground state and the $j$-th excited state of
the ancilla system respectively. If this Hamiltonian and the ancilla
system with a large damping rate are available, the encoded system
will be protected by the AQEC process while the errors in the ancilla
system will not propagate to the encoded system.

Currently, the experimental efforts mostly concentrate on the corrections of errors during gate operations, and the universal get set $\left\{\hat{H}, \hat{S}, \hat{T}, \mathrm{cPhase}\right\}$ in an error-transparent manner are feasible in experiment. However, the ultimate universal quantum computation
requires the suppression of the error rate to an arbitrarily small level
when the elementary gate operation fidelities exceed a certain threshold,
without requiring a physical resource overhead scaled exponentially.
\red{Although the fault-tolerant threshold is still lacking for the bosonic
codes, there are opportunities to further extend the bosonic codes along two directions. One is to promote the performance of the single-mode codes by increasing the mean photon number of the codewords and thus utilizing the higher-order encoding to tolerate more errors~\cite{Michael2016}. The other one is to extend the system to multiple modes by repeating the strategies used in their qubit counterpart and employing the non-local information encoding for achieving the fault tolerance~\cite{GuillaudPRX2019}.}


\begin{figure*}
\centering{}\includegraphics[width=1.8\columnwidth]{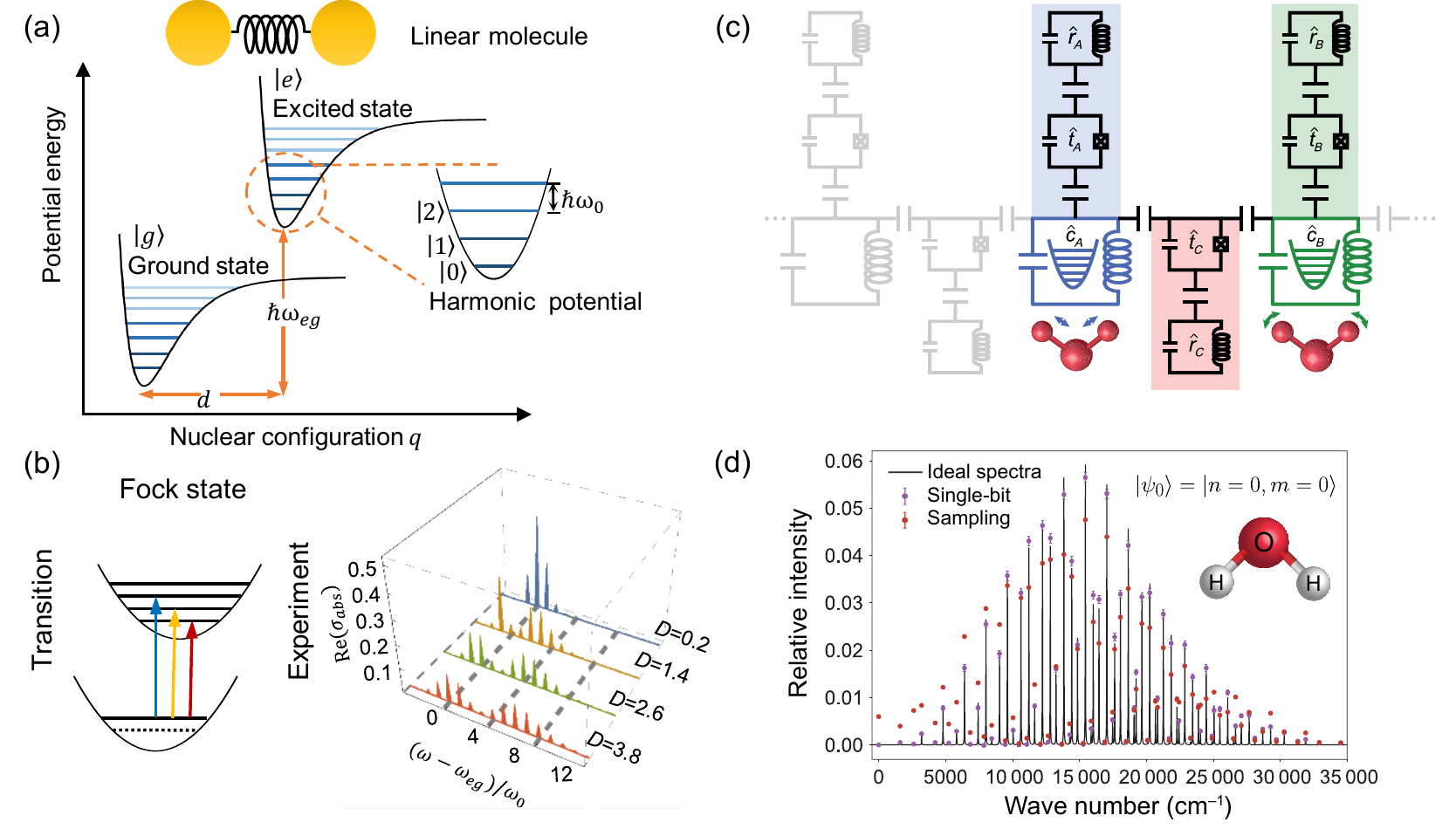}
\caption{\textbf{Quantum simulation based on bosonic modes.} (a) Principle
of a superconducting simulator for the vibronic structure of diatomic
molecules. (b) The absorption spectrum in the molecular system with
different Huang-Rhys parameters $D$, where the initial state is a
non-equilibrium Fock state. (a-b) are adapted from Ref.~\cite{Hu2018simulation}. (c) Circuit schematic of a two-mode
superconducting bosonic processor for simulating molecular vibronic
spectra and extracting Franck-Condon factors for photoelectron processes.
(d) Experimental Franck-Condon factors for photoionization of water.
(c-d) are adapted from Ref.~\cite{WangChristopher2020}.}
\label{fig:QuantumSimulation}
\end{figure*}

\subsection{Quantum communications with bosonic codes}
\label{sec:QuantumCommunications}
In a quantum network~\cite{Kimble2008,Reiserer2015}, quantum information needs to distribute among quantum \red{nodes (or modules)} via either direct quantum state transfer or quantum teleportation through shared quantum entanglement. As a result, efficient quantum state transfer between quantum nodes and on-site long-lifetime quantum memories are essential for a quantum network. \red{On one hand, photons are the most practical choice for high rate communications between distinct nodes. On the other hand, quantum information encoded by bosonic codes can be protected by QEC from local noise during storage, channel noise associated with wavepackets propagating over distances, and the insertion loss at quantum interfaces due to impedance mismatching.} Therefore, the bosonic codes are of great potential for building quantum networks, and proof-of-principle experiments of most basic quantum network components have been reported.

As sketched in Fig.~\ref{fig:QuantumCommunications}, each node can consist of a storage cavity, a readout cavity, and a transmon qubit dispersively coupled to both cavities. The stored bosonic codes in the storage cavity can be converted to a traveling wavepacket through a coherent frequency conversion between the two cavities based on two coherent drives and a four-wave mixing effect~\cite{Pfaff2017}, as shown in Fig.~\ref{fig:QuantumCommunications}a.
Connecting two nodes, quantum state transfer via a cable coupled to two
readout cavities has been demonstrated with a pitch-and-catch protocol.
Based on the binomial codes, the dominant error (single photon loss)
in the communication can be detected and corrected, and on-demand
entanglement between quantum nodes has been demonstrated~\cite{Axline2017},
as shown in Fig.~\ref{fig:QuantumCommunications}d. In a different
setup (Fig.~\ref{fig:QuantumCommunications}c), the entanglement
between quantum memories has been realized by a standing mode of a superconducting coaxial bus resonator~\cite{burkhart2020errordetected}. The bosonic mode encoding in the even parity subspace enables the tracking of photon loss events during the two-photon interference to promote the fidelity of the generated entanglement.

\red{Based on the above demonstrated components, quantum repeaters could
be realized with superconducting bosonic codes.} Besides, quantum communication can also be realized without direct interaction between nodes by \red{quantum state
teleportation}, which only requires entanglement shared between nodes,
local operation, and classical communication. \red{Equiped with quantum repeaters and quantum state teleportation,} quantum information could then be delivered over arbitrarily long distances with high fidelity through practically imperfect quantum communication channels, as required for \red{secure quantum communication on planetary scale.} In addition, distributed quantum computation could be realized also based on quantum entanglement shared between nodes. This module-based approach could avoid spurious cross-talks between components, as well as the frequency crowding in device engineering. In distributed quantum computation, teleportation-based quantum gate operations between separated quantum nodes are critical. Recently, deterministic teleportation of a CNOT gate between two nodes (both with bosonic encodings) is demonstrated~\cite{Chou2018}, as illustrated in Fig.~\ref{fig:QuantumCommunications}b (also Fig.~\ref{fig:LogicalGates}d).

The limitation of the superconducting bosonic system in building a
practical quantum network is mainly imposed by the thermal noise at
room temperature. In contrast, optical photons could transmit information
over thousands of kilometers~\cite{Ren2017}, \red{while being restricted
only} by probabilistic quantum gate operations. Therefore, the ideal microwave-to-optical transducers~\cite{Fan2018,Han2020} are required for taking advantage of both microwave and optical bosonic codes. For instance, a theoretical study predicts a high secure key rate for memory-less one-way quantum communication over long distances with cat codes~\cite{Li2017}. In addition to communications, quantum networks could also enhance the sensing or measurement by distributing correlated quantum probes. For example, higher precision could be achieved in a longer-baseline quantum telescope~\cite{GottesmanPRL2012}.

\subsection{Quantum simulations with bosonic codes}
\label{sec:QuantumSimulation}

Although the ultimate universal quantum computation is extremely challenging,
the use of noisy quantum systems in quantum simulation has attracted immediate
research interests~\cite{Houck2012}. In the NISQ era~\cite{Preskill2018}, early quantum simulations could find direct applications in exploring quantum chemistry, quantum optimization, material engineering, as well as fundamental studies of condensed matter physics and high-energy physics, and could also stimulate further research interest in the universal quantum computation. Compared with qubit arrays, the bosonic modes are indispensable in many physical models, including the boson sampling, molecular vibration, quantum Rabi model, Bose-Hubbard model, and the simulation of a non-Markovian environment. Besides, the bosonic modes could also be applied directly in analog quantum simulations.

As an example, the bosonic modes are applied to solve the vibrational
structure problem. To make accurate calculations of the vibrational structure of large systems is still very challenging for classical computers. Instead of manipulating qubits in conventional quantum simulators in the absence of natural
properties of elementary particles, bosonic simulators are competent
to establish direct correspondence between photonic cavity modes and
molecular vibrational modes. Analog quantum algorithms are capable
of simulating molecular vibrations. A proof-of-principle experiment
has demonstrated how superconducting devices can simulate the vibronic
spectra of molecules~\cite{Hu2018simulation}. The device comprises
of a transmon qubit coupled to a 3D cavity, where the two lowest energy levels of the qubit are manipulated as the electronic ground and excited
states of a molecule, while the bosonic mode of the cavity models
the nuclear vibrational motion of the molecule. By offering the vibronic structure
of diatomic molecules, the simulator can obtain the molecular spectra
for both equilibrium and non-equilibrium states. Further experimental efforts
are paid on extending the system to multiple bosonic modes. A superconducting bosonic processor that integrates two superconducting microwave cavities and three transmon qubits has been realized, where each cavity represents one vibrational mode of a triatomic molecule and the qubit-mediated coupling represents the interaction between the modes. Based on a high-fidelity single-shot photon number detection scheme that is capable of resolving up to 15 photons, the photoelectron spectra of several triatomic molecules, including $\mathrm{H_{2}O}$, $\mathrm{O_{3}}$, $\mathrm{NO_{2}}$, and $\mathrm{SO_{2}}$, are simulated~\cite{WangChristopher2020}, proving a bright future of the bosonic modes in analog quantum simulations.

\begin{figure*}
\centering{}\includegraphics[width=1.8\columnwidth]{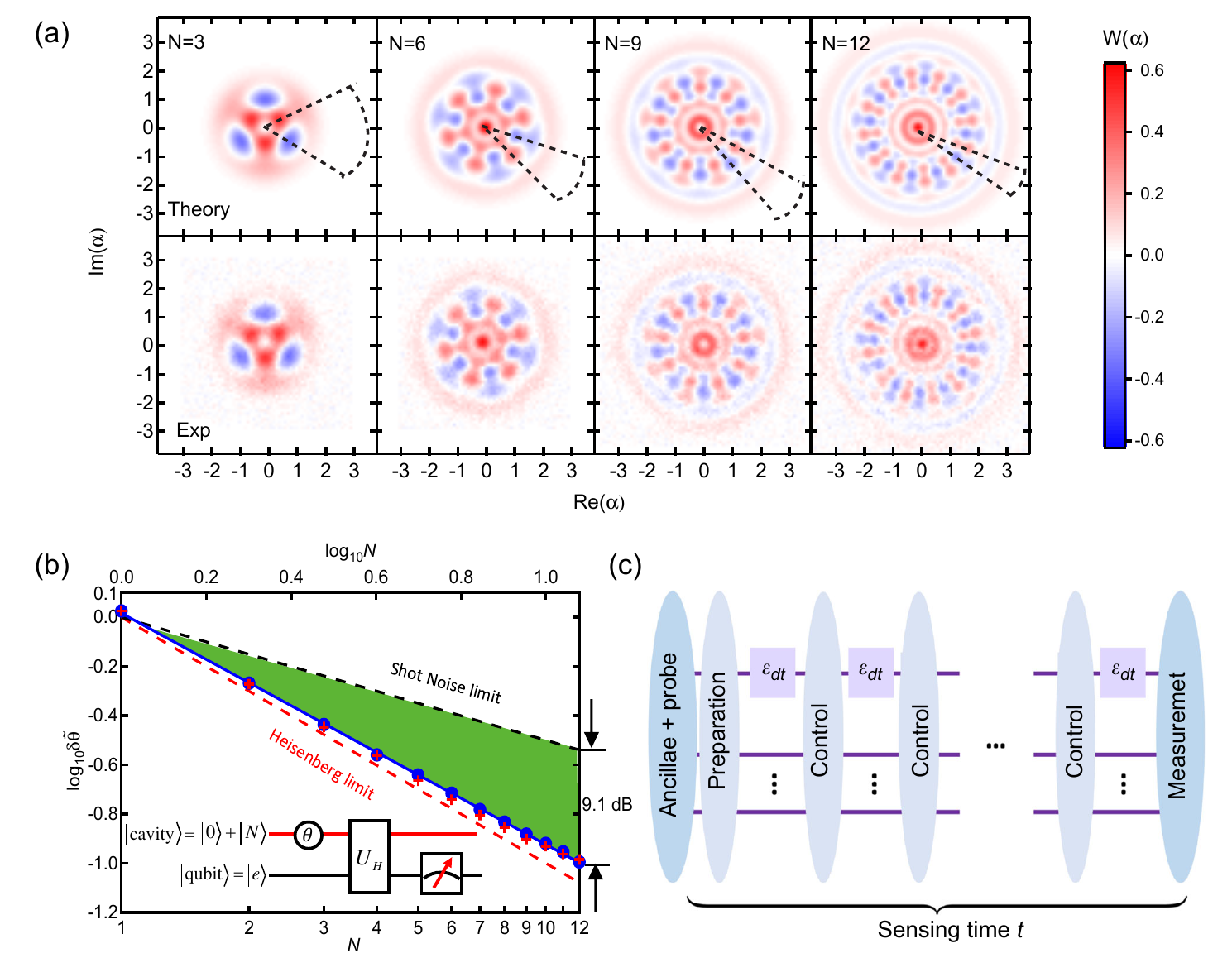}
\caption{\textbf{Quantum metrology based on bosonic modes.} (a) Theoretical and experimental Wigner functions of the maximum variance states $(|0\rangle+i|N\rangle)/\sqrt{2}$ encoded in a microwave mode.
(b) Results of optimal single-mode sensing scheme. Blue
dots are experimental results and green region represents the experimental
results that surpass the standard limit by about 9.1 dB at $N=12$.
(a-b) are adapted from Ref.~\cite{WangNC2019Heisenberg}.
(c) The QEC-enhanced metrology scheme. One probe sequentially senses
the parameter for time $t$ with quantum controls applied every $dt$.
Adapted from Ref.~\cite{Zhou2018}.}
\label{fig:QuantumMetrology}
\end{figure*}

The digital quantum simulation of topological phases is also
carried out in the superconducting bosonic system~\cite{Flurin2017}, in which the
spin-orbit coupled particles running on a lattice is efficiently simulated.
This digital simulator performs a split-step quantum walk algorithm
and directly measures the associated topological invariant by using the interference between two components of a cavity Schr{\"{o}}dinger
cat state. The direct measurement of such a quantity in solid-state
materials remains a significant challenge, owing to the non-local
nature of the topological ordering. \red{This protocol sheds light on the
simulation and characterization of complex quantum materials based on superconducting bosonic modes.}


In these previous bosonic quantum simulators, the QEC codes have not been directly put into use yet. However, we should point out that the bosonic encoding has huge potentials in digital quantum simulations. \red{Because of hardware efficiency the bosonic modes are suitable for studies of high-dimensional digital quantum simulations in the first place, and the demonstrated QEC techniques additionally allow a deeper circuit depth. The circuit depth or the fidelities could also be further improved by combining the recently proposed error mitigation method~\cite{Temme2016Error,LiPRX2017,McArdle2020}.} For a coarse estimation, assuming the imperfect logical gate operations have an operation error of $5\%$ and the bosonic codes allow an error-detection efficiency of $99\%$, we could
suppress the operation error to $1\%$. \red{For an expectation fidelity of $80\%$, we could significantly improve the circuit depth from $5$ to $20$ with a success probability of about 36\%.} Therefore, the error correction and mitigation of the bosonic codes will promote quantum simulations in the NISQ era.

\subsection{Quantum metrology with bosonic codes}
\label{sec:QuantumMetrology}
In conventional sensing and metrology applications, atom and spin
ensembles, mechanical oscillators, and microwave and optical modes are
the most used experimental systems for detecting magnetic fields,
acceleration, rotation, displacement, and distance~\cite{Degen2017RMP}. These systems could all be described or approximated by bosonic modes, and thus the
bosonic codes are of special interest for quantum-enhanced metrology.
Besides, as mentioned in Sec.~\ref{sec:QuantumCommunications} distributed quantum metrology could be realized in a quantum network~\cite{Guo2020}. \red{However, these conventional metrology techniques suffer the limited capability of nondeterministic quantum state engineering, processing, or detection.} Therefore, the superconducting systems and their hybridization with spins or mechanical
resonators provide a unique platform for realizing \red{high-performance} quantum metrology.

When estimating a parameter $\omega$ through the interaction Hamiltonian
of an oscillator as $H(\omega)=\omega H_{\mathrm{I}}$ and by preparing the oscillator mode in a coherent state with a mean photon number $N$, the precision
of the parameter estimation is limited from two aspects: the classical
shot-noise in detecting photons $\propto1/\sqrt{N}$ and the coherence
time $T_c$-limited interaction duration \red{$\propto1/T_c$}~\cite{Giovannetti2011}.
However, these limits are not as fundamental as the Heisenberg uncertainty
principle, which imposes an ultimate limit in measurement precision $\propto1/N$,
called the Heisenberg limit (HL)~\cite{Helstrom1976,Holevo1982,Giovannetti2011,Giovannetti2004,Giovannetti2006}.
By exploring the large Hilbert space of a bosonic mode, both above limitations could be resolved.

On one hand, the shot-noise could be suppressed by preparing the mode
in a quantum state that gives a maximum variance for $H_{\mathrm{I}}$. The
interferometers composed of two bosonic modes have been implemented
on various platforms by utilizing squeezed states,
number states, and Schr{\"{o}}dinger cat
states~\cite{Haroche,Polino2020}. Instead of fragile two-mode states, quantum
metrology schemes based on a single bosonic mode have also been experimentally
implemented in trapped ions and superconducting circuits~\cite{McCormick2019,WangNC2019Heisenberg}.
Especially, an enhanced sensitivity approaching the HL scaling is demonstrated~\cite{WangNC2019Heisenberg}.

On the other hand, the Hilbert space intrinsically provides redundancy to construct a QEC code subspace, which could be mapped to orthogonal
subspaces by errors and recovered back through error correction. Therefore,
the coherence time of the probing quantum state could be extended
by protecting the code subspace from environment noise via QEC~\cite{Zhou2018}. Combining QEC and universal operation on a binomial code, a recent
work demonstrates a Ramsey experiment on the QEC protected logical
qubit and shows a coherence time twice as long as that without QEC~\cite{Hu2019}.
Since Ramsey interferometry has been widely used for precision measurements,
this result reveals the potential of bosonic codes in sensing. \red{Although the QEC-enhanced quantum metrology} has attracted considerable
attention, it is still challenging for experiments. One challenge
comes from the so-called Hamiltonian-not-in-Lindblad-span (HNLS) condition
for the existence of an optimal code that can be constructed for achieving the
HL scaling~\cite{Zhou2018}. Recently, by an approximate QEC technique,
the advantage of QEC for a bosonic radiometry has been demonstrated in a superconducting circuit~\cite{Unpublished}, though the HNLS condition is not completely satisfied. This experiment indicates that the bosonic QEC has considerable potential to be explored in quantum metrology.

\section{Discussions and outlook}
\label{sec:discussions}
The bosonic codes in a superconducting quantum system hold the advantages
of hardware efficiency, large Hilbert space, and unique capability of
long-distance transfer, therefore are one of the most promising candidates
for future quantum applications. Although great potentials of the
bosonic codes have been revealed by many preliminary experimental results,
there are many challenges to be addressed in the future studies.

For short-term research, we would expect further extensions of
current bosonic systems and demonstrations of quantum advantages
brought by the bosonic codes. Even though universal fault-tolerant quantum
information processing is not available yet, the bosonic QEC technique is
beneficial for the protection of quantum information from temporal, propagation loss, and gate errors, allowing longer storage time, longer propagation distance, and deeper quantum circuit depth. So, we would expect immediate explorations of bosonic codes in quantum repeater, quantum simulation, quantum metrology, and quantum machine learning with the near-term NISQ superconducting systems~\cite{Preskill2018}. \red{At this stage, these applications could unarguably stimulate more research interests from both experimental and theoretical perspectives, which would encourage new ideas about the optimization and applications of the bosonic codes and might also even reveal new physics of the bosonic codes.}

In addition, we need to further extend the bosonic system to multiple-oscillator regimes and other bosonic oscillators. A resonator array has been demonstrated in 2D~\cite{Naik2017}, and the 3D micromachined microwave cavities could also be scalable by a multilayer integration approach~\cite{Brecht2016}. As required
for long-distance quantum communication and quantum network, high-efficiency and low-noise quantum transducers that convert the microwave signals to optical frequencies are significant. Recently, there are exciting progresses along this direction: direct and coherent transducers based on superconducting cavity electro-optics~\cite{Fan2018} and high-frequency phonon-mediated piezo-electro-optomechanical coupling~\cite{Han2020} are demonstrated, both of which avoid the MHz-frequency mechanical noise. On the other hand, the mechanical modes provide a more compact platform for high-density integration of bosonic modes. Such a hybrid phononic architecture allows the realization of multimode mechanical memory for quantum random accessing memory~\cite{Hann2019},
and is also useful in the \red{mechanical-oscillator-based force or inertial
sensing~\cite{Jacobs2017}.}

Long-term goals of universal quantum computation demand more efforts, and here we summarize the challenges from three aspects:

(i) \emph{Material and fabrication}. Superconducting hardware is the backbone of quantum information technology. Improvements of the superconducting materials and fabrication techniques are always worthwhile. Better understanding of the loss mechanisms~\cite{Krantz2019,Kjaergaard2020,Zmuidzinas2012}, such as quasi-particles, radiations, and piezo-mechanical losses would help superconducting qubit and cavity engineering. Combining sophisticated integration architecture and packaging technique that avoid frequency crowding and cross-talks with improved coherence times, fabrication yield, stability, and robustness, the bosonic codes could be scalable. To reduce the cost and suppress thermal background noise, it holds great potential to utilize high-frequency superconducting qubits and resonators at millimeter wavelengths for superconducting circuits that can work at high temperatures~\cite{Anferov2020}.

(ii) \emph{Theory}. For the ultimate goal of universal quantum computation,
there is still a lack of a clear estimation about the fault-tolerance threshold
for the bosonic codes. Other than extending the single-mode codes
to higher energies (higher mean photon number) \red{and higher-dimension encoding},
the extension of the bosonic codes to multiple modes is necessary.
\red{One possible approach is to concatenate the bosonic codes with the surface codes, i.e, the bosonic codes as the building blocks of the surface codes~\cite{Noh_2020}.} Another feasible approach is the realization of topological quantum codes in a distributed quantum network architecture~\cite{Nickerson2013},
by which the challenges due to the massive integration of cavities
in a single module to avoid cross-talks and frequency crowding problem
could be relaxed. A hardware-adaptive code could be numerically optimized according to the practical system parameters, and the studies on the interconversions between different bosonic and qubit-based codes are also needed. We might expect new fault-tolerant bosonic quantum computation architectures. Besides, efforts are needed for the applications of bosonic codes in quantum metrology, quantum simulations, and quantum networks.

(iii) \emph{Advanced quantum control techniques}. The limited quantum gate fidelity is actually the main obstacle for demonstrating high-order bosonic codes that are able to correct more errors, because the control pulse sequences would be more complicated due to the larger dimension of the Hilbert space. The fidelity losses mainly originate from three aspects, i.e. the system incoherent processes, incomplete physical model in the numerical optimization of the control parameters, and parameter errors in the experimental setup. Although these losses are determined by the hardware imperfections, advanced quantum control techniques would help. Robust quantum control could minimize the fidelity loss
against the parameter fluctuations, and a more complete physical model
by including open quantum system dynamics as well as higher-order
nonlinear interactions could be developed by a hybrid quantum-classical
approach. By introducing the recently developed machine learning control
methods, device calibration and quantum algorithms might be implemented
with higher efficiency. Additionally, as pointed out in Sec.~\ref{sec:UniversalControl}, most current studies focus on the spin-oscillator model in the strong dispersive interaction regime, however, \red{a combination of the moderate Pockel or Kerr nonlinearities with the spin-oscillator model} might extend our capability of universal quantum control, especially when extending the bosonic codes to higher mean photon numbers.

In summary, this article summarizes the recent development of bosonic
QEC codes in a superconducting quantum platform. The bosonic modes are universal in nature, and thus the demonstrations in the superconducting quantum circuits could be directly extended to optical frequencies, mechanical oscillators, and spin wave in spin ensembles. Especially, the tools demonstrated in the spin-oscillator model
could also be equipped in the spin-phonon systems based on trapped-ions
and NV centers, as well as the optical cavity QED systems. We believe
that the bosonic codes will be fruitful in both short-term and long-term
future and will play an indispensable role in quantum information
technologies.


%

\end{document}